\documentclass[fleqn,usenatbib]{mnras}

\usepackage{newtxtext,newtxmath}

\usepackage[T1]{fontenc}

\DeclareRobustCommand{\VAN}[3]{#2}
\let\VANthebibliography\thebibliography
\def\thebibliography{\DeclareRobustCommand{\VAN}[3]{##3}\VANthebibliography}

\usepackage{tabularx}
\usepackage{graphicx}	
\usepackage{amsmath}	
\usepackage{makecell}
\usepackage{multirow}
\usepackage{xspace}

\usepackage[flushleft]{threeparttable}

\newcommand{\csixtyplus}{C$_{60}^{+}$\xspace}

\title{Families and Clusters of Diffuse Interstellar Bands: a Data-Driven Correlation Analysis}

\author[Fan et al.]
{
Haoyu Fan,$^{1,2}$\thanks{E-mail: hfan58@uwo.ca}
Madison Schwartz,$^{1,2}$
Amin Farhang,$^{3,4}$
Nick L. J. Cox,$^{5}$
Pascale Ehrenfreund,$^{6}$
\newauthor
Ana Monreal-Ibero,$^{6}$
Bernard H. Foing,$^{6}$
Farid Salama,$^{7}$ 
Klay Kulik,$^{1,2}$
Heather MacIsaac,$^{1,2}$
\newauthor
Jacco Th. van Loon,$^{8}$
Jan Cami$^{1,2,9}$
\\
$^{1}$Department of Physics and Astronomy, The University of Western Ontario, London, ON N6A 3K7, Canada\\
$^{2}$The Institute for Earth and Space Exploration, The University of Western Ontario, London, ON N6A 3K7, Canada\\
$^{3}$Department of Physics, University of Tehran, North Karegar Ave, 14395-547 Tehran, Iran\\
$^{4}$School of Astronomy, Institute for Research in Fundamental Sciences, 19395-5531 Tehran, Iran\\
$^{5}$ACRI-ST, 260 Route du Pin Montard, 06904 Sophia-Antipolis, France\\
$^{6}$Leiden Observatory, Leiden University, Niels Bohrweg 2, 2333 CA Leiden, The Netherlands\\
$^{7}$NASA Ames Research Centre, Space Science \& Astrobiology Division, Moffett Field, California, USA\\
$^{8}$Lennard-Jones Laboratories, Keele University, ST5 5BG, UK\\
$^{9}$SETI Institute, 189 Bernardo Ave, Suite 100, Mountain View, CA 94043, USA\\
}

\date{Accepted XXX. Received YYY; in original form ZZZ}

\pubyear{2021}

\begin{document}
\label{firstpage}
\pagerange{\pageref{firstpage}--\pageref{lastpage}}
\maketitle

\begin{abstract}
More than 500 diffuse interstellar bands (DIBs) have been observed in astronomical spectra, and their signatures and correlations in different environments have been studied over the past decades to reveal clues about the nature of the carriers. We compare the equivalent widths of the DIBs, normalized to the amount of reddening, $E_\textrm{B-V}$, to search for anti-correlated DIB pairs using a data sample containing 54 DIBs measured in 25 sight lines. This data sample covers most of the strong and commonly detected DIBs in the optical region, and the sight lines probe a variety of ISM conditions. We find that 12.9\% of the DIB pairs are anti-correlated, and the lowest Pearson correlation coefficient is $r_{\rm norm}\sim -0.7$. We revisit correlation-based DIB families and are able to reproduce the assignments of such families for the well-studied DIBs by applying hierarchical agglomerative and $k$-means clustering algorithms. We visualize the \textit{dissimilarities} between DIBs, represented by 1 - $r_\textrm{norm}$, using multi-dimensional scaling (MDS). With this representation, we find that the DIBs form a rather continuous sequence, which implies that some properties of the DIB carriers are changing gradually following this sequence. We also find at that least two factors are needed to properly explain the dissimilarities between DIBs. While the first factor may be interpreted as related to the ionization properties of the DIB carriers, a physical interpretation of the second factor is less clear and may be related to how DIB carriers interact with surrounding interstellar material.
\end{abstract}

\begin{keywords}
ISM: lines and bands -- ISM: molecules -- dust, extinction
\end{keywords}

\section{Introduction}\label{sec:introduction}
The Diffuse Interstellar Bands (DIBs) are a set of absorption features that represent a century-long mystery regarding the interstellar medium (ISM). The first DIBs, $\lambda\lambda$5780 and 5797\footnote{We follow the convention that the DIBs are referred to by their approximate central wavelength expressed in \AA}, were noted towards $\zeta$ Per by \cite{1922LicOB..10..146H}, and now more than 500 such absorption features have been catalogued in the optical region \citep{2008ApJ...680.1256H, 2009ApJ...705...32H, APOcat}, and tens more in the near-infrared \citep[e.g.][]{1990Natur.346..729J, 2014A&A...569A.117C, 2015ApJ...800..137H, 2016ApJ...821...42H}. The substructures within several DIB profiles strongly suggest a molecular origin \citep[e.g.][]{1995MNRAS.277L..41S, 1996A&A...307L..25E, 1998ApJ...495..941K, 2001ApJ...561..272W, 2004ApJ...611L.113C}, yet their specific carriers remain unknown, maybe except the two near-infrared DIBs at $\lambda\lambda$9577 and 9633 and three weaker DIBs that have been assigned to \csixtyplus -- a finding that is supported by an impressive array of observational and experimental studies (\citealp{1994Natur.369..296F, 1997A&A...317L..59F, 2015Natur.523..322C, 2015ApJ...812L...8W, 2016ApJ...831..130W, 2017ApJ...843...56W, 2017ApJ...843L...2C, 2017ApJ...846..168S, Lallement:EDIBLES2, 2019ApJ...875L..28C}; see \citealp{Harold:C60plusreview} for a review). We note that this assignment was recently challenged by \citet{2021AJ....161..127G} who reported that the two DIBs $\lambda\lambda$9577 and 9633 are poorly correlated -- they report a Pearson correlation coefficient $r=0.32$. A thorough review of this claim however finds the opposite, that the $\lambda\lambda$9577 and 9633 DIBs do in fact correlate very well ($r\sim0.9$; Schlarmann et al., in press), thus further supporting this identification.

To guide and support laboratory efforts to identify more DIB carriers, astronomers perform analyses of astronomical observations to provide constraints on the properties of DIB carriers and thus narrowing down the candidates to be examined. Such efforts often include correlation studies, where the DIB strengths, most often represented by their equivalent widths (EWs, or denoted as $W$(DIBs)), are compared to ISM parameters such as $E_\textrm{B-V}$, column densities of various species ($N$(X)), or to other DIBs  \citep[e.g.][]{Herbig1993, Cami:DIBcorrelations, Friedman2011, Vos2011, behavior}. It has been recognized long ago that many DIBs respond to changing physical conditions (termed the "DIB behaviour", typically observed by changes in the EW), but not necessarily in the same way \citep{1974ApJ...194..313S, 1991MNRAS.252..234A, 1994A&A...281..517J,Cami:DIBcorrelations, Vos2011, 2014IAUS..297...13S}. The classical example in this context is illustrated by comparing the $\lambda\lambda$5780 and 5797 DIBs in the sight lines toward $\zeta$~Oph (HD~149757; $E_\textrm{B-V}$=0.28) and $\sigma$~Sco (HD~147165, $E_\textrm{B-V}$=0.34). The sight line of $\zeta$~Oph crosses a cloud interior and thus probes an environment that is shielded from UV irradiation. On the other hand, the sight line of $\sigma$~Sco represents a more exposed environment. In both sight lines, the EW of the $\lambda$5797 DIB is similar ($\sim$35~m\AA), while $W$(5780) increases from $\sim$73~m\AA\ in $\zeta$~Oph to $\sim$240~m\AA\ in $\sigma$~Sco. Clearly, the $\lambda$5797 DIB carrier is somewhat indifferent to the changing UV exposure, whereas the $\lambda$5780 is very sensitive to it. This could indicate different photo-chemical properties of the DIB carriers themselves but could also be due to more indirect effects where the UV radiation dissociates molecular hydrogen which in turn then affects the chemical network of DIB carriers \citep{1993MNRAS.262..831W}.

The changing response to environment then also leads to the idea of DIB "families". Members of the same family show a similar response to environmental factors and thus also exhibit good mutual correlations, whereas members of different families show a much poorer correlation \citep{1987ApJ...312..860K, 1989A&A...218..216W, Cami:DIBcorrelations, 2003MNRAS.338..990W}. Strong correlations between two DIBs from the same family could also indicate the same or related carriers. While there is no general agreement about precisely which DIBs belong to a specific family, some connections are well established. For instance, the $\lambda\lambda$5797, 6379 and 6613 DIBs show similar behaviour, favouring less exposed ($\zeta$-type) environments, and have been grouped into a family by various authors. In this paper, we will refer to this family of DIBs as the $\zeta$-DIBs. The DIBs $\lambda\lambda$5780 and 6284 (along with several others) on the other hand thrive in exposed ($\sigma$-type) regions \citep{2015MNRAS.452.3629L, PCA_Cami, 2018PASP..130g1001K, 2020A&A...637A..74O}. They can also be considered as members of a family that we will call $\sigma$-DIBs for what follows. In this context, another group of the so-called C$_2$-DIBs can be seen as a third family, whose members trace dense and molecular regions of the ISM cloud \citep{C2DIB}. 

While many studies have focused on correlations, anti-correlations between DIBs could also be of particular interest. Such DIB pairs could indicate that their carriers are the start and end products of the same physical or chemical processes. For example, if one DIB would be carried by a neutral species and another DIB by its cation, one would expect the strength of one DIB to decrease as the strength of the other increases. Similar arguments of course hold for other processes such as hydrogenation. A key issue that plagues correlation studies however is that different sight lines typically represent different amounts of interstellar material. Since more interstellar material in general also implies higher column densities for individual species, the EWs of DIBs thus always have some positive correlation with each other \citep[e.g.][]{2015MNRAS.454.4013B, 2016A&A...585A..12B}. This effect tends to hide anti-correlations. But by comparing the normalized EWs of DIBs ($W$(DIB)/$E_\textrm{B-V}$), \cite{Cami:DIBcorrelations} were among the first to identify several anti-correlated DIB pairs within a small sight line sample. 

Motivated by more robustly confirming the existence of anti-correlated DIB pairs, we revisit the topic of DIB (anti-)correlations as well as DIB families. We include more DIBs than most previous analyses to obtain a more general picture of DIB correlations, rather than focusing on a handful of well-studied DIBs. This paper is organized as follows. Section \ref{sec:Data} describes the data we use and the selection criteria of our target DIBs. We search for anti-correlated DIB pairs in Section \ref{sec:correlation}, and sort the DIBs into groups in Section \ref{sec:clustering} according to their mutual correlations. Section \ref{sec:MDS} contains our efforts to visualize the DIB correlations, and the implications of our findings are discussed in Section \ref{sec:discussion}. Finally we summarize this work in Section \ref{sec:summary}.

\section{Data and DIB Selection}\label{sec:Data}
\begin{table*}
	\centering
	\caption{Properties of the 54 target DIBs in this work}
	\label{table: DIB properties}
	\begin{threeparttable}
	\begin{tabular}{cccccc|cccccc} 
		\hline
		\hline
		Label & Wave. & FWHM & \makecell{No. of\\Mea.} & \makecell{Avg. \\ Nor. EW} & Comments & Index & Wavelength & FWHM & \makecell{No. of\\Mea.} & \makecell{Avg. \\ Nor. EW} & Comments \\
		\AA & \AA & \AA &  & m\AA /mag &  & \AA & \AA & \AA &  & m\AA /mag \\
		\hline
		4429 & 4429.33 & 24.13 & 10 & 2005.95 & Broad DIB\tnote{1} & 6270\tnote{2} & 6269.88 & 1.48 & 24 & 104.64 & \\ 
		4501 & 4501.51 & 2.53 & 17 & 73.84 &  & 6284 & 6284.05 & 4.49 & 24 & 958.09 & \\
		4726 & 4726.98 & 2.76 & 21 & 168.87 & C$_2$ DIB & 6324 & 6324.91 & 0.74 & 20 & 10.00 & \\
		4762 & 4762.44 & 1.94 & 20 & 48.81 & & 6330 & 6330.03 & 0.73 & 21 & 10.11 & \\
		4963 & 4963.92 & 0.68 & 25 & 29.20 & C$_2$ DIB & 6353 & 6353.31 & 1.66 & 20 & 26.77 & \\
		4984 & 4984.78 & 0.51 & 18 & 16.11 & C$_2$ DIB & 6362 & 6362.26 & 1.61 & 20 & 18.76 & \\
		5418 & 5418.87 & 0.75 & 20 & 22.01 & C$_2$ DIB & 6367 & 6367.30 & 0.52 & 21 & 11.50 & \\
		5512 & 5512.68 & 0.54 & 20 & 14.85 & C$_2$ DIB & 6376 & 6376.14 & 0.76 & 23 & 32.59 & \\
		5545 & 5545.08 & 0.84 & 20 & 25.02 & & 6377 & 6377.07 & 0.57 & 22 & 11.34 & \\
		5546 & 5546.46 & 0.68 & 21 & 12.38 & C$_2$ DIB & 6379 & 6379.25 & 0.64 & 23 & 79.12 & \\
		5705 & 5705.12 & 2.68 & 20 & 90.65 & & 6397 & 6397.04 & 1.27 & 20 & 23.85 & \\
		5766 & 5766.16 & 0.76 & 22 & 16.31 & & 6439 & 6439.51 & 0.82 & 23 & 18.31 & \\
		5780 & 5780.67 & 2.09 & 17 & 398.84 & & 6445 & 6445.30 & 0.60 & 20 & 23.56 & \\
		5793 & 5793.24 & 0.96 & 20 & 14.32 & C$_2$ DIB & 6449 & 6449.27 & 0.94 & 21 & 19.77 & \\
		5797 & 5797.18 & 0.89 & 23 & 140.63 & & 6520 & 6520.74 & 1.00 & 21 & 25.07 & \\
		5828 & 5828.50 & 0.78 & 20 & 10.96 & & 6553 & 6553.88 & 0.51 & 21 & 11.04 & \\
		5849 & 5849.82 & 0.83 & 24 & 50.88 & & 6613 & 6613.74 & 1.05 & 25 & 185.14 & \\
		5923 & 5923.51 & 0.74 & 21 & 16.34 & & 6622 & 6622.84 & 0.58 & 21 & 9.67 & \\
		6065 & 6065.32 & 0.64 & 21 & 12.19 & & 6660 & 6660.67 & 0.63 & 23 & 39.22 & \\
		6089 & 6089.85 & 0.58 & 22 & 18.74 & & 6699 & 6699.28 & 0.67 & 22 & 24.66 & \\
		6108 & 6108.06 & 0.49 & 20 & 8.34 & & 6702 & 6702.07 & 0.74 & 22 & 10.05 & \\
		6116 & 6116.80 & 0.87 & 20 & 10.56 & & 6729 & 6729.22 & 0.29 & 21 & 9.16 & C$_2$ DIB \\
		6185 & 6185.79 & 0.48 & 22 & 6.38 &  & 6993 & 6993.12 & 0.77 & 21 & 64.72 & \\
		6196 & 6195.99 & 0.51 & 24 & 47.06 & & 7224 & 7224.16 & 1.12 & 19 & 162.17 & \\
		6203 & 6203.58 & 1.63 & 25 & 157.31 & & 7367 & 7367.08 & 0.64 & 20 & 11.50 & \\
		6212 & 6211.69 & 0.62 & 20 & 8.95 & & 7559 & 7559.43 & 0.91 & 20 & 14.17 & \\
		6234 & 6234.01 & 0.65 & 20 & 16.93 & & 7562 & 7562.16 & 1.55 & 20 & 51.96 & \\
		\hline
	\end{tabular}
  \begin{tablenotes}
    \item[1] See \cite{broadDIB}
    \item[2] This DIB was divided into three separate DIBs in \cite{APOcat} to reflect the structures within its profile. We remeasured the entire feature as one DIB in this work to follow the convention of most DIB studies.
  \end{tablenotes}
  \end{threeparttable}
\end{table*}

We base our work on the DIB measurements from \cite{APOcat}, which is part of the DIB survey project carried out at the Apache Point Observatory (APO) and the University of Chicago. We refer the reader to the paper series ``Studies of the Diffuse Interstellar Bands'' and related publications \citep[i.e.][]{2001ApJ...559L..49M, C2DIB, 2008ApJ...680.1256H, 2009ApJ...705...32H, 6196-6613, Friedman2011, 2013ApJ...773...41D, 2014ApJ...792..106W, behavior, APOcat} for more details on the data and measurement techniques than outlined here. 

The DIB measurements are made with R $\sim$ 38,000 spectra towards 25 medium to highly reddened targets ($E_\textrm{B-V}$ between 0.31 and 3.31 $mag$). The spectral types of the background stars are between O6 and A5, and the sight lines cover a great variety of ISM conditions as characterized by their $f_{H2}$ values. To identify the presence of stellar lines that may compromise the DIB measurements, each of these target stars is paired with a standard star with similar spectral type but very low reddening. Table \ref{table: sightline info} summarizes the information on the target stars and their corresponding standard star, and a full version of the same table can be found in \cite{APOcat}. Telluric lines are removed from the raw data by fitting a template spectrum based on air mass, and a telluric reference spectrum is displayed during the DIB measuring process to provide guidance on possible residuals from the correction. 

Direct integration is used to measure the EWs of DIBs without assuming any specific profile, and the uncertainties are estimated based on the signal-to-noise ratio and the width of the profile. Since the spectra data are normalized by the data-reduction pipeline, only a local continuum is needed around the target DIB. Consistent measuring techniques, especially regarding the selections of continuum regions and integration limits, are kept for each DIB to all sight lines. Such effort ensures a uniformly-made data set with great self-consistency. We flag defects such as contamination from adjacent stellar/telluric lines or large uncertainty in the continuum level, and exclude such measurements in the analysis. We also include $E_\textrm{B-V}$ and column densities of some molecular species in our study. The sources of these data are described in \cite{behavior}.

The selection of target DIBs in this work is based on the number of measurements available among the 25 sight lines, and we require each of the target DIBs to have no more than five excluded measurements or upper limits (non-detection). This is to ensure that many environments can be considered in the DIB correlations and that the resulting correlations are robust. However, we made exceptions for a few DIBs of particular interest, such as the broad DIB $\lambda$4429 and the strong DIB $\lambda$5780. In total we include 54 DIBs in our correlation analysis as summarized in Table \ref{table: DIB properties}. These DIBs cover most of the strong and/or well-studied optical DIBs in the literature \citep[e.g.][]{Cami:DIBcorrelations, Cox2006, Vos2011, Friedman2011, 2013ApJ...774...72K}.

\section{Searching for Anti-correlated DIBs}\label{sec:correlation}
As is the case for any interstellar material, DIB carriers generally would accumulate over distance. Hence their EWs always have a positive correlation with each other to a certain degree, especially when some of the most heavily reddened sight lines are included in the data sample \citep{Friedman2011}. To reveal possible anti-correlations among DIBs, we choose to follow the approach outlined in \cite{Cami:DIBcorrelations} and work with the EWs of DIBs normalized by the reddening (i.e. using $W$(DIB)/$E_\textrm{B-V}$). When discussing Pearson correlation coefficients using this normalized EW, we will denote them with $r_\textrm{norm}$ to differentiate them from the ``regular'' correlations using $W$(DIBs) that we will denote with $r_\textrm{reg}$. The total extinction $A_\textrm{V}$ also scales with the gas column densities along the sight line, and has been favoured in some works to normalize $W$(DIBs) \citep[e.g.][]{2018A&A...620A..52R}. While identifying the best normalizer of DIB strengths would be a worthy future project, it is beyond the scope of the current paper, and we found similar results as those reported in the following sections when analyze the correlations of $W$(DIB)/$A_\textrm{V}$.

\begin{figure*}
	\includegraphics[width=16cm]{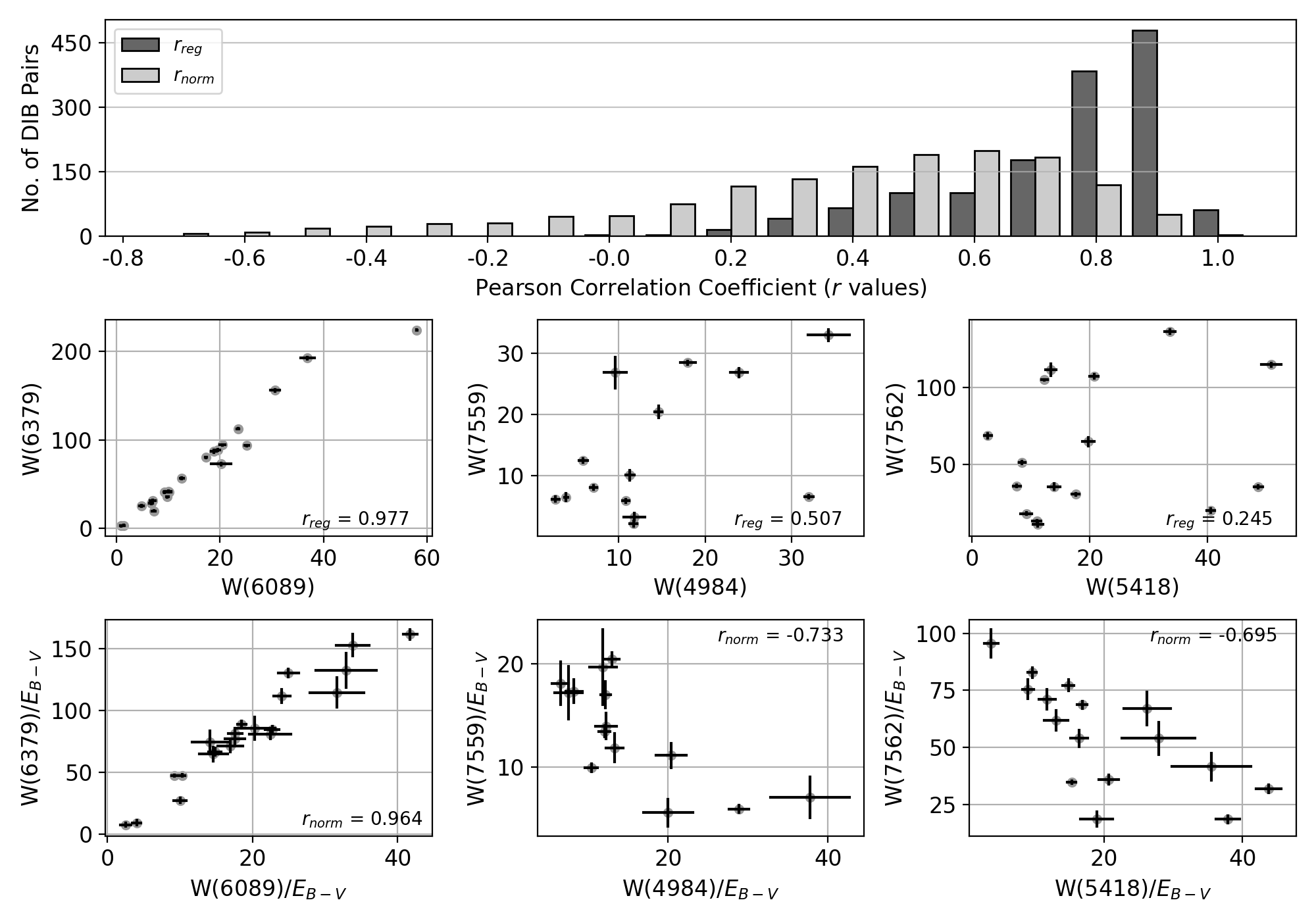}
    \caption{Comparison between $r_\textrm{reg}$ and $r_\textrm{norm}$. Upper panel: histogram of $r_\textrm{reg}$ and $r_\textrm{norm}$ values. We find $\sim$1/7 of the DIB pairs have negative $r_\textrm{norm}$ values, and the most anti-correlated DIB pairs have $r_\textrm{norm}\sim -0.7$. Middle and bottom panels: scatter plots of $W$(DIB) and $W$(DIB)/$E_\textrm{B-V}$ correlations between DIB pairs $\lambda$6089 vs $\lambda$6379, $\lambda$4984 vs $\lambda$7559, and $\lambda$5418 vs $\lambda$7562. The first pair has the highest $r_\textrm{norm}$ value, while the second and third are the most anti-correlated DIB pairs in our data sample. The units for $W$(DIB) and $W$(DIB)/$E_\textrm{B-V}$ are m\AA\ and m\AA$\cdot$ mag$^{-1}$, respectively.}
    \label{fig: regular vs normalized}
\end{figure*}

Figure \ref{fig: regular vs normalized} provides some comparisons between $r_\textrm{norm}$ and $r_\textrm{reg}$. The $r_\textrm{norm}$ value is in most cases smaller than $r_\textrm{reg}$ value of the same DIB pair. The highest $r_\textrm{norm}$ value is between $\lambda\lambda$6089 and 6379 with $r_\textrm{norm}$ = 0.964, and there are 22 DIB pairs with $r_\textrm{norm} \geq $ 0.9. Many of these DIB pairs are known to be well-correlated when the comparison is between their EWs \citep[e.g.][]{6196-6613, 2021MNRAS.507.5236S}, such as $\lambda$6196 vs $\lambda$6613 ($r_\textrm{reg}$ = 0.980, $r_\textrm{norm}$ = 0.952) and $\lambda$6203 vs $\lambda$6284 ($r_\textrm{reg}$ = 0.991, $r_\textrm{norm}$ = 0.936).

While all $r_\textrm{reg}$ values are positive, we find 184 DIB pairs have negative $r_\textrm{norm}$ values. They take up 12.9\% of the total possible combinations among the 54 target DIBs. The most anti-correlated DIB pairs, $\lambda$4984 vs $\lambda$7559, and $\lambda$5418 vs $\lambda$7562, have $r_\textrm{norm} \sim -0.7$, and their scatter plots are presented in the middle and lower panel of Figure \ref{fig: regular vs normalized}. As indicated by the plotting axes, $W$(DIBs)/$E_\textrm{B-V}$ still vary over a factor of 5 to 10 among the target sight lines. This reflects the impact of the environmental factors on DIBs, and how DIBs may be used to trace such differences \citep{Vos2011, 2013ApJ...774...72K, 2013A&A...550A.108V, 2015MNRAS.454.4013B, behavior}.

\begin{figure}
	\includegraphics[width=\columnwidth]{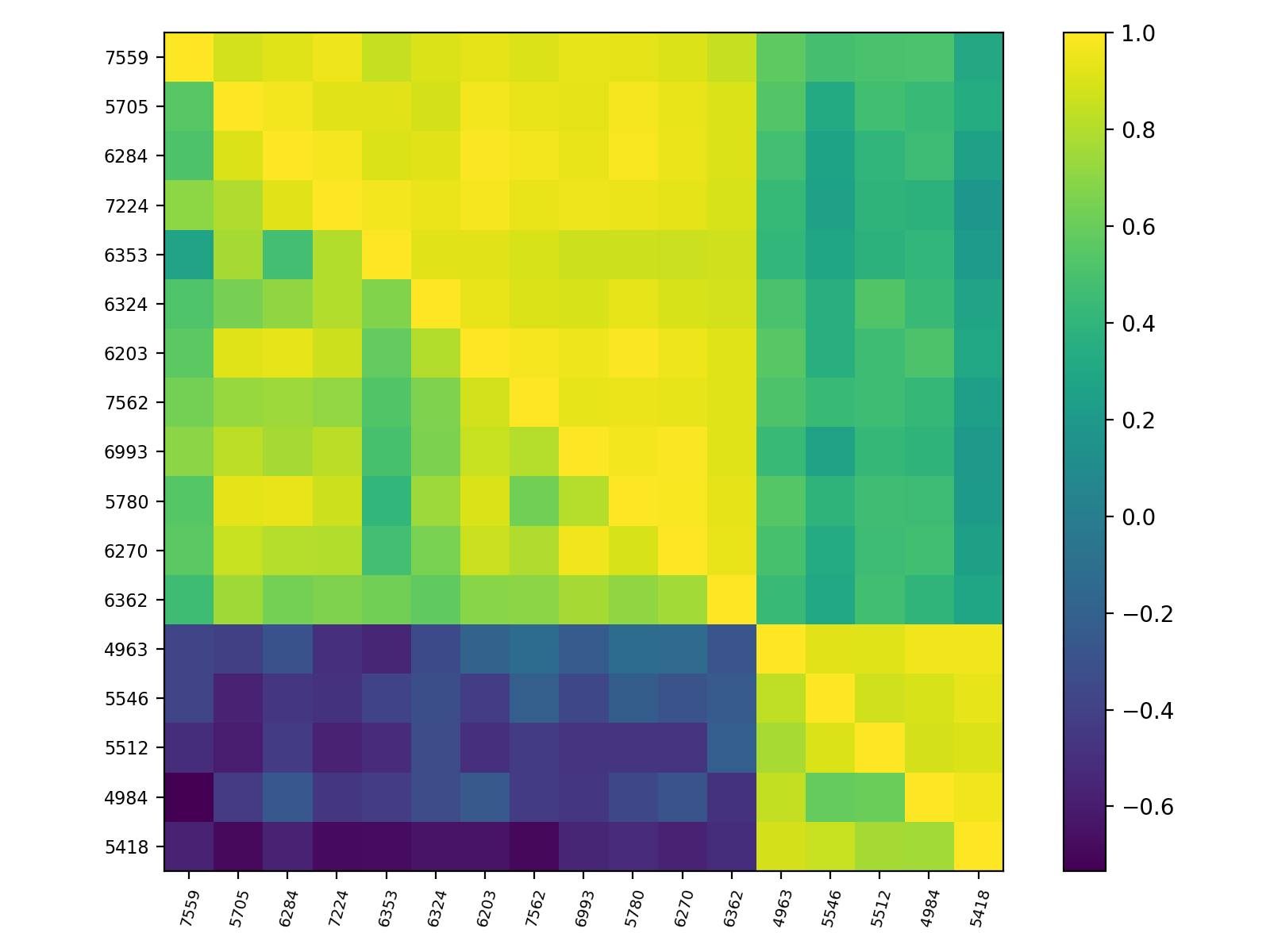}
    \caption{Heat map for the Pearson correlation coefficients among the 17 DIBs involved in the 40 DIB pairs with $r_\textrm{norm} < -0.5$. Note the figure is asymmetrical, where the lower triangle is for the $r_\textrm{norm}$ values and the upper triangle is for the $r_\textrm{reg}$ values. The DIBs are sorted by the sequence in Section \ref{sec:MDS} and two groups can be identified. They are respectively parts of the $\sigma$-type and C$_2$ DIB groups (Section \ref{sec:clustering}), and all cross-group comparisons yield negative $r_\textrm{norm}$ values. A full-scale heat map for all target DIBs of this work is presented in the Appendix \ref{appendix: page-wide plots}.}
    \label{fig: smallheatmap}
\end{figure}

While the \textit{r} values offer a good estimation on the similarity between two DIBs, the exact values depend on the composition of the sight line sample and the uncertainties in the measurements. There are 40 DIB pairs with $r_\textrm{norm} \leq -0.5$ and 17 unique DIBs are involved in these pairs. We show their correlation coefficients in the heat map of Figure \ref{fig: smallheatmap}, and two DIB groups emerge. The first group consists of DIBs $\lambda\lambda$5705, 5780, 6203, 6270, 6284, 6324, 6353, 6362, 6993, 7224, 7559, and 7562, and many of them are known to be DIBs that remain prominent under strong radiation (i.e. in $\sigma$-type environments). The second group contains $\lambda\lambda$4963, 4984, 5418, 5512, and 5546. They are all C$_2$-DIBs that trace denser regions of interstellar clouds and display different dependencies on environmental factors like radiation and density \citep{C2DIB, behavior}. This result suggests that DIB clusters or families can be identified according to their mutual correlations \citep[e.g.][]{Cami:DIBcorrelations, 2003MNRAS.338..990W, 2020A&A...637A..74O}. That is, members of the same group share more similarities and thus have better correlations, while DIBs from different groups have reduced $r$ values due to their different preferences on the environments.

\section{Clustering of DIBs}\label{sec:clustering}
The apparent clustering of the C$_2$-DIBs in the previous section encouraged us to study clustering of the DIBs more closely based on their correlation coefficients. We first constructed a $54\times54$ matrix of $r_\textrm{norm}$ values\footnote{We use the $r_\textrm{norm}$ values since this work is originally motivated by the search for anti-correlated DIB pairs. We will show in Appendix \ref{appendix:different_r} that the general picture of DIB correlations does not change if the analysis is based on $r_\textrm{reg}$ values.}. Each row of this matrix is an array of 54 $r_\textrm{norm}$ values between 1.0 and -1.0 and can be interpreted as a coordinate of a parameter space. DIB families can be identified when a group of DIBs are located closely in this parametric space, i.e. when they are are mutually well-correlated and have similar $r_\textrm{norm}$ values with other DIBs outside the group. 

\begin{figure*}
	\resizebox{\hsize}{!}{\includegraphics[width=\textwidth]{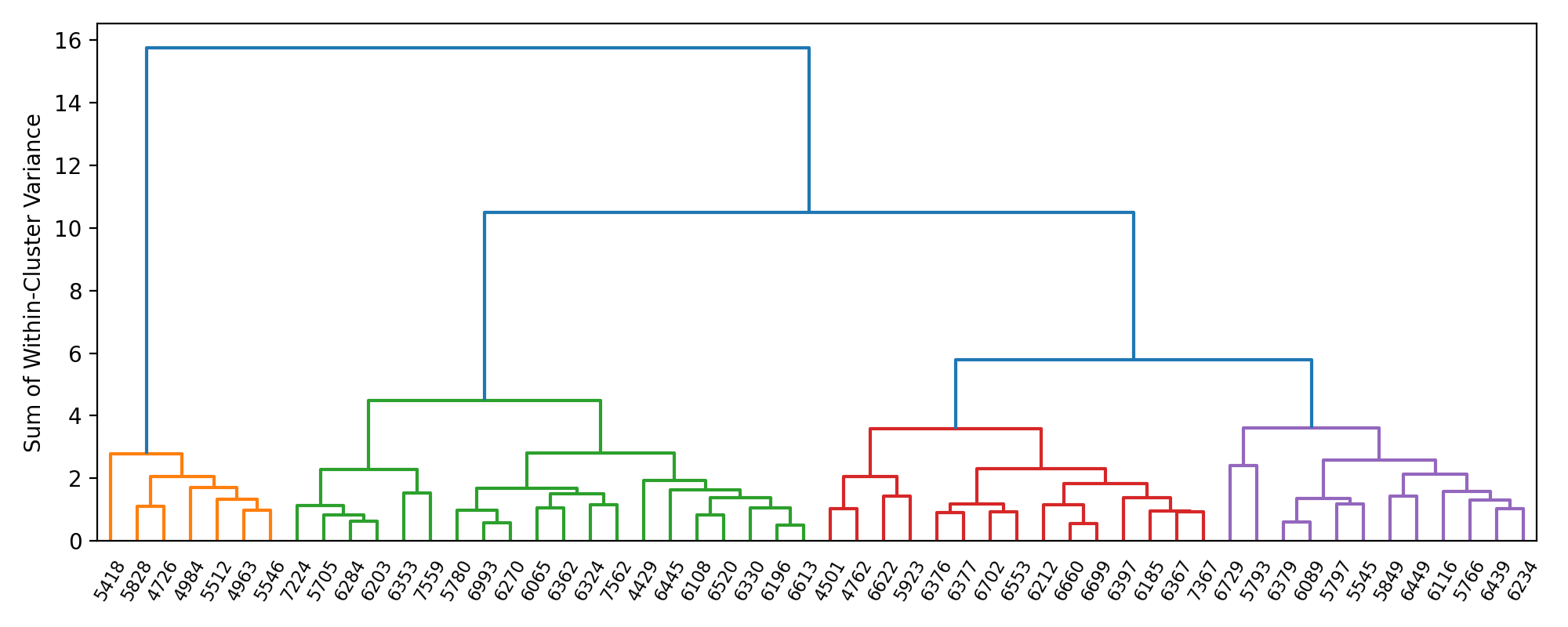}}
    \caption{Dendrogram for the hierarchical agglomerative clustering analysis. The 54 DIBs are arranged along the x-axis and gradually merged into the root on top of the plot. The horizontal bars indicate which clusters/DIBs are being merged during each iteration, and the sum of within-cluster variance after the merge. We choose to keep four clusters that correspond to the $\sigma$-type (green), $\zeta$-type (purple), C$_2$ DIBs (orange), plus an ``intermediate'' (red) group whose properties is between the $\sigma-$ and $\zeta$-types of DIBs.}
    \label{fig: Dendrogram}
\end{figure*}

As a first attempt to cluster the target DIBs, we applied hierarchical agglomerative clustering (HAC\footnote{not to be confused with hydrogenated amorphous carbon that has been proposed as a DIB carrier}) to the $r_\textrm{norm}$ matrix. With this method, the algorithm initially takes each DIB as a singleton cluster. During each iteration, two clusters are merged in such a way that the sum of within-cluster variance is kept minimal after they are merged (Ward's minimum variance method \citep{WardMethod}). This process is repeated until all DIBs are merged into a single cluster, and is similar to the approach by \cite{2015MNRAS.451..332B} except we use a different ``linkage'' function that determines which clusters to merge. 

This process is graphically represented in Figure~\ref{fig: Dendrogram} as a dendrogram or ``tree diagram''. In the figure, the DIBs are sorted along the x-axis and gradually merged into nodes and finally the root on the top, and DIBs connected by a lower node share more similarities. The criterion to decide where precisely to cut off the tree (vertically) will then decide on the number of branches (clusters) to be kept. This criterion is subjective. However, as will be discussed in the next section, the number of groups is not very important since the DIBs in fact exhibit a rather continuous sequence. We thus choose to keep four clusters, indicated by different colours in Figure \ref{fig: Dendrogram}). This clustering accounts for the three known DIB families discussed above (the $\lambda$5797 family, the $\lambda$5780 family and the C$_2$-DIBs) plus a possible unidentified group. This results in the following four clusters:

\begin{enumerate}
    \item The $\sigma-$DIB group (green in Fig.~\ref{fig: Dendrogram}), containing $\lambda\lambda$4429, 5705, 5780, 6065, 6108, 6196, 6203, 6270, 6284, 6324, 6330, 6353, 6362, 6445, 6520, 6613, 6993, 7224, 7559, and 7562;
    \item The intermediate DIB group (red), containing $\lambda\lambda$4501, 4762, 5923, 6185, 6212, 6367, 6376, 6377, 6397, 6553, 6622, 6660, 6699, 6702, and 7367;
    \item The $\zeta-$DIB group (purple), containing $\lambda\lambda$5545, 5766, 5793, 5797, 5849, 6089, 6116, 6234, 6379, 6439, 6449, and 6729;
    \item The C$_2$ DIB group (orange), containing $\lambda\lambda$4726, 4963, 4984, 5418, 5512, 5546, and 5828.
\end{enumerate}

Unlike the classical approach where several DIBs are grouped solely for having good correlations, the HAC algorithm also considers whether they are less correlated with other DIBs to the same degree. This difference should be subtle since two perfectly-correlated DIBs would always have the same \textit{r} values with a third DIB. Our results are fully consistent with the literature for the well-known DIBs, such as for the typical $\sigma$-type DIBs $\lambda\lambda$5705, 5780, and 6284, as well as $\zeta$-type DIB $\lambda\lambda$5797 and 6379 \citep{2015MNRAS.452.3629L, PCA_Cami, 2018PASP..130g1001K, 2020A&A...637A..74O, 2021AJ....161..127G}. We also have eight C$_2$ DIBs from the original reference \citep[i.e.][]{C2DIB} and find six of them in the C$_2$ DIB group, while the two exceptions are assigned to the $\zeta$-type group whose members also prefer shielded environments. Our clustering efforts also expand the knowledge to some less-often targeted DIBs. For example, DIBs $\lambda\lambda$5849 and 6379 are a factor of two stronger in the shielded sight line of BD+63$^\circ$ 1964 compared to HD~183143 \citep{1997A&A...318L..28E}, and our analysis confirms them as $\zeta$-type DIBs like $\lambda$5797. Lastly, a new ``intermediate'' DIB group is introduced. We will show in Section \ref{sec:MDS} that members of this group have properties between the $\sigma$-type and $\zeta$-type DIBs, and together they form a rather continuous spectrum of DIB behaviour.

We also adopt $k$-means clustering, another widely-used clustering algorithm, to test the robustness of the grouping result. The $k$-means clustering algorithm sorts all data points into \textit{k} groups so that: a) the centre of each group is given by the average coordinate of its members, and b) each data point is closer to its own group centre than to other group centres. The $k$-means clustering results are overall quite similar to our HAC results. Ten DIBs however are assigned to a different ``adjacent'' group compared to the HAC results. Indeed, the seven DIBs in the $\lambda\lambda$6196 and 6613 sub-branch of Figure \ref{fig: Dendrogram} are assigned to the intermediate group rather than $\sigma$-group; the $\lambda$6397 DIB is assigned to the $\sigma$-group rather than intermediate group; the $\lambda$6553 DIB is assigned to the $\zeta$-group rather than intermediate group; and the $\lambda$6729 DIB, recognized as a C$_2$ DIB in \cite{C2DIB}, to C$_2$ group rather than $\zeta$-group. As will be shown in the next section, these DIBs are mostly located around the ``junction regions'' between clusters. Their membership of specific DIB families is thus somewhat ambiguous, and the uncertainty in their membership would not change the overall picture of DIB behaviour that emerges from this clustering.


\section{Multi-Dimensional Scaling analysis}\label{sec:MDS}
While clustering algorithms sort the DIBs into groups, they provide limited information on any possible linkages between these groups. In this section we use a Multi-Dimensional Scaling (MDS) analysis to visualize the $r_\textrm{norm}$ matrix and provide a general picture on the similarities/dissimilarities among our target DIBs. We also tested other dimensional reduction and data visualization algorithms such as t-Distributed Stochastic Neighbor Embedding \citep[tSNE, see][]{van2008visualizing} and UMAP \citep{2018arXivUMAP}, and both methods produce very similar results.

The MDS algorithm maps a set of $N$ data points onto an abstract \textit{M}-dimensional Cartesian space (with $M < N$). The only input that MDS requires is an \textit{N}-by-\textit{N} \textit{dissimilarity matrix} that contains a measure for pairwise distances among the \textit{N} observations. Somewhat akin to a Principal Component Analysis (PCA), the algorithm then maps these $N$ points onto the new \textit{M}-dimensional space in such a way that these pairwise distances are preserved as much as possible. This is done by minimizing the \textit{stress} function:

\begin{equation}
\label{equ: stress}
        \textrm{stress} = \sqrt{\frac
        {\sum_{i,j}^{}(d_{ij}-\hat{d}_{ij})^2}
        {\sum_{i,j}^{}d_{ij}^2}
        }
\end{equation}

\noindent Here $d_{ij}$ is the observed distance from the input matrix, and $\hat{d}_{ij}$ is the distance between points in the new $M$-dimensional space mapped by the algorithm. The details of how the algorithm optimizes this mapping can be found in \citet{MDS1}. An classical example of MDS is to feed it pairwise distances between cities and let MDS reconstruct a map from those distances. 

Since we are using Pearson correlation coefficients that measure the similarity between two DIBs, we use $1 - r_\textrm{norm}$ as input values for the dissimilarity matrix. In this way the dissimilarity between two DIBs is minimized to zero when they are perfectly correlated, and maximized to 2.0 when they are perfectly anti-correlated. We also include correlations with the column densities of H$_2$, CH, C$_2$, and CN in the analysis -- i.e. we expanded our dissimilarity matrix to the size of 58$\times$58 that includes correlations with these column densities. Their $r_\textrm{norm}$ values are calculated normalized to the $E(B-V)$ as is the case for the DIBs, and the resulting $1 - r_\textrm{norm}$ is added to the dissimilarity matrix.

\begin{figure}
	\resizebox{.9\hsize}{!}{\includegraphics[width=\columnwidth]{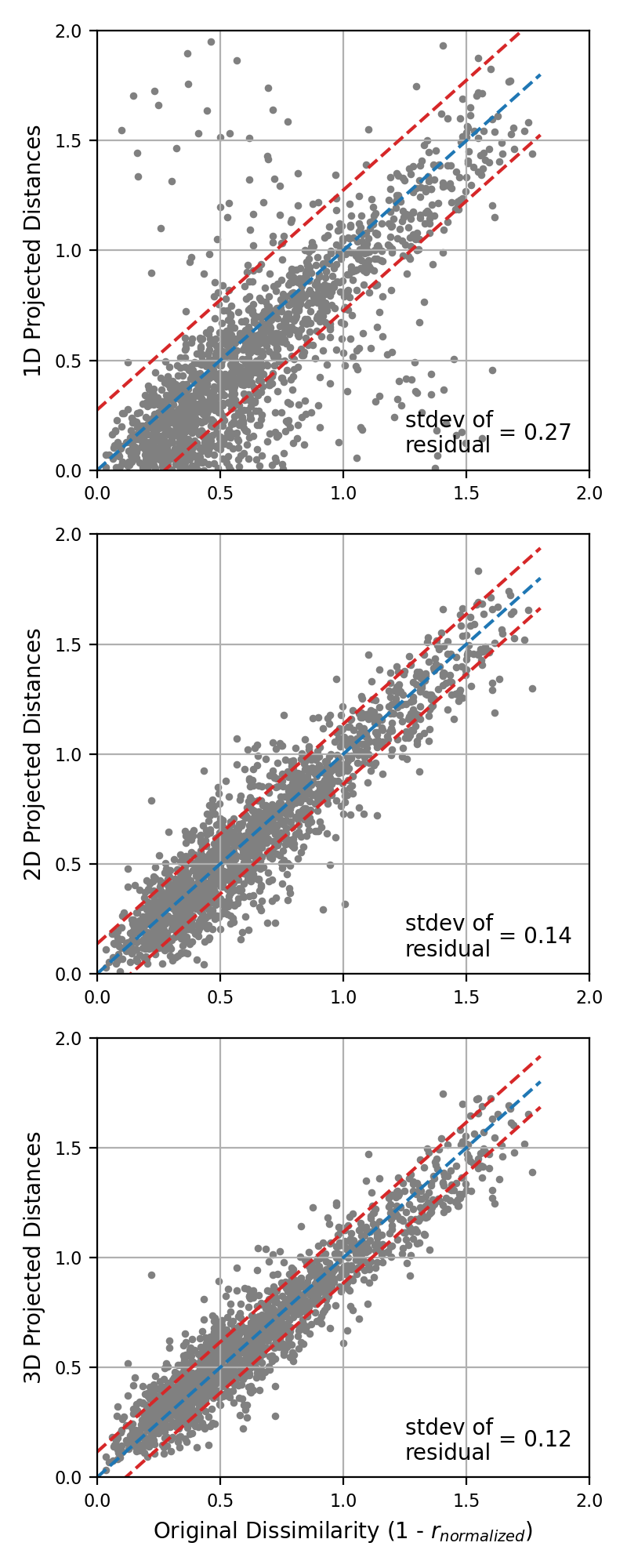}}
    \caption{Comparing the results of the 1D-, 2D-, and 3D-MDS analyses. Each panel compares the input $d$ = 1 - $r_\textrm{norm}$ values (abscissa) to the MDS-mapped distances (ordinate). The blue dashed lines represent the ideal case $Y = X$ and the red dashed lines illustrate the standard deviation of the residuals.}
    \label{fig: MDS_distance}
\end{figure}

Including more dimensions (i.e. the higher $M$ is) would always reduce the stress (Equation \ref{equ: stress}) and bring better agreement between the input dissimilarities and the distances in the new $M$-dimensional space. However it is equally important to use a small number of dimensions so the results are easier to be interpreted. For the visualization purposes, it is most common for MDS to map the dissimilarity matrix to a 1D-, 2D-, or 3D-space so the result can be demonstrated as scatter plots \citep{MDS2}. In Figure \ref{fig: MDS_distance} we examine the results of 1D-, 2D-, and 3D-MDS by comparing the pairwise input dissimilarity [$d_{ij}$] = [1 - $r_{ij}$] to the pairwise distances [$\hat{d}_{ij}$] in the new $M$-dimensional space. If the MDS algorithm would have preserved all distances, the comparison should yield $Y = X$ but we find considerable scatter in the top panel that shows the output of the 1D-MDS. Mapping the DIBs along a single axis is thus not sufficient to fully explain their dissimilarities, and an additional mapping axis is required. The scatter is greatly reduced after adding a second projection axis, and including a third axis provides little improvement (Figure \ref{fig: MDS_distance} middle and bottom panels). We will thus adopt the result of the 2D-MDS analysis on our dissimilarity matrix for the remainder of this paper. 

\begin{figure}
	\includegraphics[width=\columnwidth]{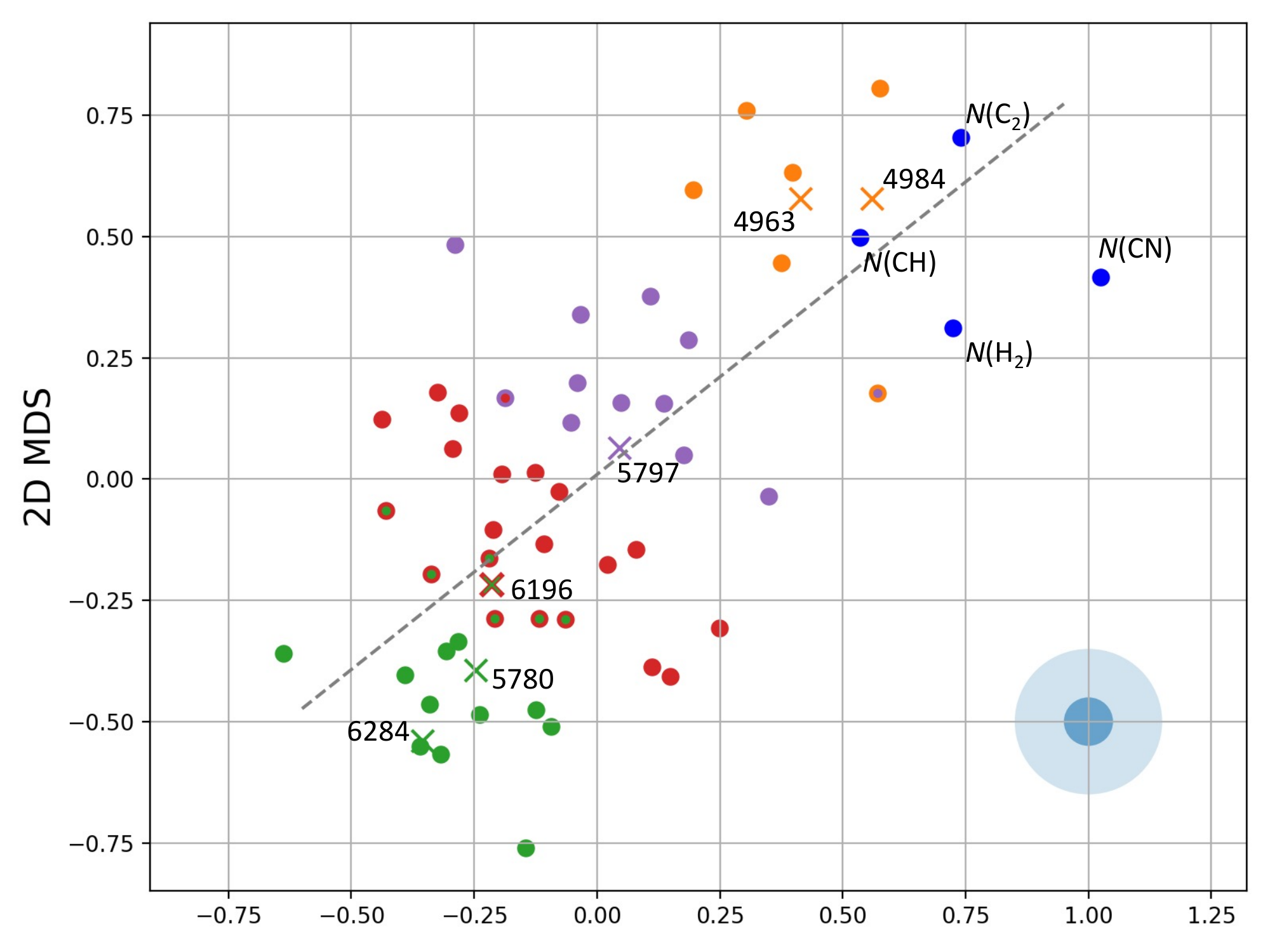}
	\caption{Scatter plot for the 2D-MDS results, where each point represents a DIB (or other tracer) and the distance between them represents their $1 - r_\textrm{normalized}$ dissimilarity. The plotting axes are abstract coordinates in the new 2D space and do not necessarily represent a physical quantity -- hence they are not labelled. We follow the same colour code as in Figure \ref{fig: Dendrogram} for DIB groups, except for the ten DIBs assigned to different groups by the HAC and $k$-means clustering algorithms. For these DIBs, the face (inner) colour represents the HAC result and the edge (outer) colour represents the $k$-means result, and they are located around the transition regions between clusters. DIBs $\lambda\lambda$6284, 5780, 6196, 5797, 4963, and 4984 are highlighted, and we also include $N$(H$_2$), $N$(CH), $N$(C$_2$), and $N$(CN) for comparison (blue dots). The dashed line is a least-squares fit representing a straight line through all DIB points. The transparent blue circles in the lower-right corner have radii of 0.05 (dark blue) and 0.15 units (light blue), and thus correspond to $r_\textrm{norm}$ values of 0.95 and 0.85 respectively. Thus, points that are separated by the radius of the light blue circle have a mutual correlation coefficient of 0.85. We find a rather smooth transition from one DIB group to the next especially among the non-C$_2$ DIBs. The molecular species are all located around the C$_2$ DIBs and they seem to form a somewhat separate cluster from the non-C$_2$ DIBs. We also provide an enlarged version of this plot in Figure \ref{fig: MDS_result_all_label}, where all data points have been labelled.}
    \label{fig: MDS_result}
\end{figure}

Figure \ref{fig: MDS_result} presents the 2D-MDS result, i.e. the locations of the normalized parameters ($W$(DIBs)/$E_\textrm{B-V}$ or $N$(Xs)/$E_\textrm{B-V}$) in the new 2-dimensional space, using the same colours for different DIB groups as in Figure \ref{fig: Dendrogram}. The ten DIBs assigned to different groups by the HAC and $k$-means clustering algorithms are plotted with different edge and face colours. Note the MDS analysis focuses on distances between points, and that distances would not change if the projection coordinate is rotated, flipped, or translated. Thus, the plotting axes of Figure \ref{fig: MDS_result} do not necessarily have a physical meaning. We focus on the general layout and trend of the projected points.

The clustering result in Section \ref{sec:clustering} remains valid in the MDS analysis. From the lower left to the upper right of Figure \ref{fig: MDS_result}, we find in a roughly linear manner of the $\sigma$-type, intermediate, $\zeta$-type, and finally the C$_2$ DIBs and the molecular species. This trend agrees with the general knowledge on how DIBs react to environmental factors especially regarding the radiation field: $\sigma$-DIBs like $\lambda\lambda$5780 and 6284 are much more prominent in radiative environments than in shielded environments, whereas $\zeta$-DIBs like $\lambda$5797 are strong in shielded environments as well, and the C$_2$ DIBs trace denser, more shielded regions than other DIBs. Such diversity in DIB behaviour is often associated with the ionization potentials of their carriers \citep{Cami:DIBcorrelations, 1997A&A...327.1215S}, although other mechanism like hydrogenation, dehydrogenation, and depletion may be influencing DIBs and other ISM species alike \citep[e.g.][]{1994Sci...265..209C, 2007ApJ...669..378J, 2007ApJ...669..401J, 2010MNRAS.404.1321W, 2014ApJ...797L..30Z, behavior}. We also find DIBs assigned to different groups by the HAC and $k$-means clustering algorithms to be mostly located in the inter-cluster regions. These DIBs may not be the ``typical members'' of any of the DIB groups described earlier, making their assignments more difficult. 

\cite{behavior} proposed a sequence of eight DIBs based on how their strength ratios change with the $f_\textrm{H2}$ value of the sight line. In order of favouring decreasing radiation and increasing shielding, the sequence goes $\lambda\lambda$6284 and 5780, then $\lambda\lambda$6196 and 6613, then $\lambda$5797, and finally the C$_2$ DIBs $\lambda\lambda$4727, 4963, and 4984. This is fully consistent with the more expanded trend observed in Figure \ref{fig: MDS_result}. To extract this sequence in our data, we perform a linear fit to the DIBs (i.e. excluding all $N$(Xs)) and project all points to this best-fit line. The observed sequence is as follows: $\lambda\lambda$7559, 5705, 6284, 7224, 6353, 6324, 6203, 7562, 6993, 5780, 6270, 6362, 6065, 6520, 6445, 6108, 6196, 6613, 6330, 5923, 4429, 6367, 6376, 6212, 6622, 7367, 6377, 4501, 6702, 6397, 6699, 6660, 6553, 6185, 4762, 6379, 5797, 5793, 6234, 6089, 6439, 5766, 5545, 6116, 6449, 5849, 5828, 6729, 4726, 4963, 5546, 5512, $N$(CH), $N$(H$_2$), 4984, 5418, $N$(C$_2$), and $N$(CN).

\section{Discussion}\label{sec:discussion}

\subsection{DIB Families}\label{sec:family}
DIB families have been discussed in various publications as the result of correlation analysis and the observed changes in their band strength ratios \citep[e.g.][]{1987ApJ...312..860K, Cami:DIBcorrelations, 1999A&A...351..680M,2015MNRAS.452.3629L}. By definition, DIBs from the same family demonstrate similar behaviour under varying ISM conditions. They are thus mutually well-correlated and have relatively constant strength ratios, and their carriers are expected to share certain properties such as ionization potentials.

The effort of DIB classification has been carried out for strong DIBs since their measurements are more accessible. Some well defined DIB families, like the $\sigma$-type, $\zeta$-type, and the C$_2$ DIBs have been described in Section \ref{sec:clustering}, and such classification echos studies on DIB profiles \citep[e.g.][]{1987ApJ...319..436J, 2003MNRAS.338..990W, 2003MNRAS.345..365G}. For example, some $\sigma$-type DIBs like $\lambda\lambda$5780 and 6284 have broad and smooth profiles, while $\zeta$-type DIB $\lambda$5797 (along with many other DIBs) have narrow profile and clear substructures, indicating gas-phase molecules as their carriers \citep{1994Natur.369..296F, 1995MNRAS.277L..41S, 1996A&A...307L..25E}.

On the other hand, the classification of certain DIBs can be ambiguous and the result varies among analyses. This issue firstly results from the uncertainties in correlation analysis, especially when weak DIBs are involved \citep[e.g.][]{Cami:DIBcorrelations, 2018PASP..130g1001K}. The \textit{r} values observed are dependent on factors such as the sample of sight lines, data quality, and the selected measuring method. It is thus very hard to compare \textit{r} values across different analyses in a qualitative manner. The assignments can still be difficult for some strong DIBs with good measurements and well-defined correlation coefficients. For example, DIBs $\lambda\lambda$6196 and 6613 are known for their close-to-perfect correlation \citep{Cami:DIBcorrelations, 1999A&A...351..680M, 2002A&A...384..215G, 6196-6613, 2016A&A...585A..12B}. But since they are equally well-correlated with both typical $\sigma$-type DIB $\lambda$6284 and $\zeta$-type DIB $\lambda$5797 \citep[e.g.][]{Friedman2011, behavior}, which DIB family should they be assigned to?

\subsubsection{$\sigma$-type, intermediate, and $\zeta$-type DIBs}\label{sec:nonC2DIBs}
By including most of the strong DIBs in the optical region, our analysis finds rather continuously distributed data points in Figure \ref{fig: MDS_result}, especially among the non-C$_2$ DIBs (i.e., the $\sigma$-type, intermediate, and $\zeta$-type groups). For most of these DIBs, several other DIBs can be found within a $\sim$ 0.15 radius, and sometimes from a different DIB family. 

This continuous trend among the non-C$_2$ DIBs goes against sorting them into several distinguishable clusters. While the behaviour of typical $\sigma$- and $\zeta$-type DIBs can be very different under different ISM conditions, there are many DIBs between them and the transition is gradual and without clear boundaries. The membership of certain DIBs to specific clusters can be thus ambiguous, such as the ten DIBs assigned to different groups by the HAC and $k$-means clustering algorithms (Section \ref{sec:clustering}). In Figure \ref{fig: MDS_result} all these DIBs are located at the junctions of the neighbouring groups, whereas the terms $\sigma$- and $\zeta$-type DIBs may only be applied to some of the most representative DIBs like $\lambda\lambda$5780, 5797 and 6284. 

DIB correlations reflect the similarities between their behaviour under different ISM conditions and thus potentially the underlying properties of their carriers. The continuous trend among non-C$_2$ DIBs suggests progressive changes in the response to their environments, due to e.g. gradually changing ionization potentials or maybe molecular sizes of their carriers.

\subsubsection{The C$_2$ DIBs}\label{sec:C2DIBs}
The C$_2$ DIBs are firstly introduced in \cite{C2DIB} and described as ``a class of weak, narrow bands whose normalized equivalent widths $W$(X)/$W$(6196) are well correlated specifically with $N$(C$_2$)/$E_\textrm{B-V}$ via power laws''. However, despite seemingly suggested by the name, many C$_2$ DIBs are in fact better correlated with $E_\textrm{B-V}$ than with $N$(C$_2$) \citep[e.g.][]{2006A&A...447..589G, EDIBLESIII_C2}. We note that some of the strongest C$_2$ DIBs are observed towards HD~37061 and HD~37903 \citep[albeit at greatly reduced strengths; see ][]{APOcat}. However, these sight lines do not show any evidence for C$_2$ absorption (Fan et al., in preparation). We thus emphasize that the detection of the C$_2$ DIBs does not depend on the prior existence of the C$_2$ molecules.

Despite a list of C$_2$ DIBs, \cite{C2DIB} does not provide a quantitative examination to check if a new DIB belongs to the family. The membership of the C$_2$ DIB family thus varies among publications when different definitions are adopted \citep[e.g.][]{2006A&A...447..589G, EDIBLESIII_C2}. In this work, six of the eight C$_2$ DIBs identified as such by \cite{C2DIB} are grouped together when using the HAC method and seven of them when performing $k$-means clustering. The remaining one or two C$_2$-DIBs in both cases are assigned to the $\zeta$-type group. This seems acceptable given the loose definition of C$_2$ DIBs, the uncertainties in the correlation coefficients, and the fact that the $\zeta$-type DIBs also trace denser regions than other non-C$_2$ DIBs. At the same time, we also identify the $\lambda$5828 DIB (not targeted in \cite{C2DIB}) as a promising new member of the C$_2$ DIB family, since both our clustering methods as well as the MDS analysis put it squarely in the same group as the other C$_2$-DIBs.

The C$_2$ DIBs are known to differ from non-C$_2$ DIBs especially regarding the ``skin effect''. This phenomenon refers to the reduced EWs of certain DIBs (relative to $E_\textrm{B-V}$) when the sight line passes through denser regions of the ISM cloud \citep{1966ApJ...144..921W, 1975ApJ...196..489S, 1984ApJ...283...98M, 1995ARA&A..33...19H}. It is best explained if those DIB carriers are more abundant in the outer layers (``skin'') of interstellar clouds, whereas the denser internal regions contribute little to the column density of DIB material. A survey of C$_2$ and C$_3$ (Fan et al., in preparation) in the EDIBLES \citep{2017A&A...606A..76C} data set finds that the EWs of the C$_2$ DIBs are in fact indifferent to C$_2$ and C$_3$ detection in the sight lines. Thus, the C$_2$ DIBs are even less sensitive to the skin effect and may thus trace denser regions than the non-C$_2$ DIBs \citep{C2DIB, behavior}. But since the C$_2$ DIBs are neither enhanced in sight lines with C$_2$, it is likely that the C$_2$ DIBs are tracing less dense regions in the ISM clouds than the C$_2$ molecules.

The current analysis suggests that the C$_2$ DIBs are really separated from the non-C$_2$ DIBs. In Figure \ref{fig: Dendrogram}, they are the two top branches of the DIB tree. In Figure \ref{fig: MDS_result}, there seems to be a boundary between the C$_2$ and $\zeta$-type DIBs, despite a continuous trend among the non-C$_2$ DIBs and that our analysis includes most of the strong optical DIBs. And finally in Figure \ref{fig: big_heat_map} that presents the \textit{r} values between all target DIBs, the non-C$_2$ DIBs appear to form an extension of the continuous progression, while there is a noticeable gap between the C$_2$ and the non-C$_2$ DIBs, especially for the $r_\textrm{reg}$ values. This gap is even more pronounced when carrying out the MDS analysis using the non-normalized correlation coefficients (i.e. based on $r_\textrm{reg}$; see Fig.~\ref{fig: MDS_r_regular}). The C$_2$ DIBs may thus arise from a very different family of molecules than the non-C$_2$ DIBs.

\subsection{Anti-Correlations}
As discussed in Section \ref{sec:correlation} the most robust anti-correlations we identify are all between a $\sigma$-type DIB and a C$_2$ DIB (see also Figure \ref{fig: smallheatmap}). The most anti-correlated DIB pairs have $r_\textrm{norm} \sim -0.7$ and are far away from a perfect anti-correlation. This lack of perfect anti-correlation is less likely to be the result of the uncertainties introduced in the normalization process, since we are able to identify plenty of DIB pairs at $r_\textrm{norm} \sim 0.9$ level. 

This lack of a perfect anti-correlation makes sense if we consider the implications from the point of view of DIB environmental behaviour. For several DIBs, it has been documented that there is a ``rise and fall'' of their strengths when comparing it to indicators for the exposure to radiation \citep[see e.g.][]{Cami:DIBcorrelations, 1997A&A...327.1215S}. \citet{behavior} demonstrate that the strengths of the $\sigma$-type DIBs such as $\lambda\lambda$5780 and 6284 demonstrate a clear $\Lambda$-shaped behaviour when compared to the mass fraction of molecular hydrogen $f_\textrm{H2}$: their $W$(DIBs)/$E_\textrm{B-V}$ peaks at a ``sweet spot'' of $f_\textrm{H2} \sim 0.2$ and decreases towards the low $f_\textrm{H2}$ end due to radiation, and towards the high $f_\textrm{H2}$ end due to the skin-effect. Given the tight mutual correlations among the $\sigma$-type DIBs, we can expect similar $\Lambda$-shaped behaviours for the other members. Thus, in order for a DIB to have a perfect anti-correlation with a $\sigma$-type DIB, it should exhibit a ``V-shaped'' behaviour with $f_\textrm{H2}$. In that case the hypothesized DIB carrier must thrive under the most exposed and shielded environment at the same time which seems not plausible.

The above discussion only applies to ideal condition with infinite sensitivities. In reality, the strengths of the C$_2$ DIBs decrease dramatically towards the low $f_\textrm{H2}$ end since they are more sensitive to the presence of radiation field. By dropping below the detection limit in low $f_\textrm{H2}$ sight lines, a C$_2$ DIB may demonstrate a portion of the required ``V-shaped'' behaviour, but the degree of anti-correlation is dependent on the composition of the sight line sample. For example, the sight lines involved in the most anti-correlated DIB pairs $\lambda$4984 vs $\lambda$7559 and $\lambda$5418 vs $\lambda$7562 all have $f_\textrm{H2} > 0.2$. In this $f_\textrm{H2}$ region the strengths of the $\sigma$-type DIBs start to decrease while $W$(C$_2$ DIBs)/$E_\textrm{B-V}$ remain roughly constant but with large scatter \citep{behavior}. The anti-correlation between the $\sigma$-type and C$_2$ DIBs thus reflects how they demonstrate different behaviours in sight lines with medium to large $f_\textrm{H2}$ values.

Since we have targeted most of the strong and commonly-detected DIBs in the optical region, it seems safe to conclude that we cannot expect perfect anti-correlations from these DIBs. On the other hand, the near infrared DIBs may demonstrate quite different behaviour than the optical DIBs \citep{2014A&A...569A.117C, 2015ApJ...800..137H, 2016ApJ...821...42H}, but we are not able to target them in our spectral data. Future projects would also benefit from including more sight lines and targeting weaker DIBs. This would also help to identify more close-to-perfect positive correlations, since molecules are expected to a few strong features along with more well-correlated weaker spectral signatures.

\subsection{Factors Governing DIB Behaviour}
\label{sec:factors}
In Section \ref{sec:MDS} and Figure \ref{fig: MDS_distance}, we find that at least two projection axes are required to properly reproduce the $1 - r_\textrm{norm}$ dissimilarities among DIBs. These two projection axes may correspond to two underlying physical factors that set the degree of correlations between DIBs.

Using a Principal Component Analysis (PCA), \citet{PCA_Cami} find that four principal components together determine $\sim 93 \%$ of the variations in the observed DIB strengths and sight line parameters in a sample of single-cloud sight lines. The first and most dominant factor is well-traced by $W$(5797) and is interpreted as the total amount of DIB-producing material along the sight line. To first order, this factor would then determine the approximate strengths of all DIBs, and be the main cause for all DIBs showing some degree of mutual correlation. The actual \textit{r} values are then determined by how much the data points deviate from this basic linear relationship in response to the physical conditions.

The second of the four factors is best traced by the $W$(5780)/$W$(5797) ratio, which is often used as a measure of UV exposure of the sight line. The radiation field can influence the behaviour of interstellar species via photo-ionization and photo-dissociation, and the presence of a strong radiation field is often associated with lower abundances of molecular species and reduced $W$(DIBs) along the sight line \citep[see e.g.][]{1977ApJ...216..291S, 1995ARA&A..33...19H, 2001ApJS..133..345W, Cox2006, Friedman2011, Vos2011}. However, it is important to realize that the $W$(5780)/$W$(5797) ratio does not trace UV exposure directly, but rather the ratio of UV exposure to the density. This is perhaps best illustrated by the sight line toward star number 46 of IC 62 that penetrates a high density photodissociation region (PDR) with $n$ > $10^4\ \textrm{cm}^{-3}$ \citep{2020MNRAS.492.5853L}. The UV radiation in this PDR is 150 times stronger than in the typical diffuse environments \citep{2018A&A...619A.170A}, leaving hydrogen in almost purely atomic form. Given the very intense radiation field one would expect most of the DIBs to have greatly reduced strengths or vanish as seen in the sight lines toward the Orion Trapezium stars \citep{behavior}. However, \cite{2020MNRAS.492.5853L} find that the normalized EWs and strength ratios of the DIBs are comparable to a regular interstellar low-$f_\textrm{H2}$ sight line. This observation thus suggests that different DIB behaviour (and thus the $W$(5780)/$W$(5797) ratio) is related to the H I/H$_2$ ratio (which in itself is determined by the radiation field and the density) rather than to the radiation field directly. This should be kept in mind for the discussion that follows. 

It is common for DIB strength ratios to vary among sight lines harbouring different environmental factors especially regarding the intensity of UV radiation. \cite{behavior} propose a sequence of eight DIBs to account for such variations \citep[see also][]{2013A&A...550A.108V}, and we find a more extended sequence in Figure \ref{fig: MDS_result}. This sequence follows the transitions between DIB families and acts as the first projection axis of our MDS analysis. In the classical radiation-shielding picture of DIB behaviours, this sequence could reflect the ionization potentials of DIB carriers. In this picture, carriers of the C$_2$ and $\zeta$-type DIBs have lower ionization potential and require more shielding, while $\sigma$-type DIBs arise from molecules of higher ionization potentials that are able to survive in more exposed environments \citep{Cami:DIBcorrelations, 1997A&A...327.1215S, 2019NatAs...3..922F}. The DIB families are formed among DIB carriers with approximate ionization potentials that would trace similar environment and thus develop good correlations.

This, of course, is a simplified picture and in reality the photo-ionization process depends on the ionization parameter $G_0 / n$ and thus the density (which has been assumed be constant in many cases), and other factors like ionization cross-sections and temperature may also contribute. One should also consider the possible role of hydrogenation and dehydrogenation that depend on the density of hydrogen and the intensity of UV radiation field  \citep[e.g.][]{2000A&A...363L...5V}. Large organic molecules, especially those with more than 50 carbon atoms, are expected to be stable and remain well-hydrogenated in typical interstellar radiation fields \citep[e.g.][]{1996A&A...305..616A, 2021A&A...650A.193O}.

Additional factors influencing the observed $W$(DIBs)/$E_\textrm{B-V}$ and thus their correlations have been proposed in the literature. For example, anomalously weak DIBs are detected towards bright stars in the LMC and SMC, and have been attributed to lower metallicities \citep{2006A&A...451..973C, 2006ApJS..165..138W, 2015MNRAS.454.4013B}. In this case, the imperfect DIB correlations may be partly due to the variable metallicity of the local ISM \citep{2021Natur.597..206D, 2021ApJS..252...22Z}. By comparing DIBs to known interstellar atomic and molecular species, \cite{behavior} find that mechanisms such as depletion onto dust grains or chemical reactions of various sorts may be required to explain the decreased $W$(DIBs)/$E_\textrm{B-V}$ in denser regions. The PCA analysis by \cite{PCA_Cami} suggests dust-related factors, such as dust-to-gas ratio or grain size along the sight line, are affecting the strength of DIBs. Many of these hypotheses involve the interaction between DIB carriers or their precursors and other species or particles, but should play a minor role compared to the radiation field. The second axis would then reflect related properties of DIB carriers that lead to different behaviour in the same environment, such as the reaction rate of the key process(es).

On the other hand, it is also possible that the second axis reflects the non-linear effect in correlation coefficients and their uncertainties \citep{Cami:DIBcorrelations}. But this would not change our findings that DIB carriers respond to the radiation field in a relatively continuous manner. These DIBs may trace certain environments, especially the diffuse atomic gas regions \citep{2013A&A...550A.108V, 2015MNRAS.454.4013B}. In principle, the strength ratios between any two DIBs that are sufficiently apart in the sequence can be used to characterize the radiation field, similar to the $W$(5780)/$W$(5780) ratio and the $\sigma - \zeta$ effect \citep{1987ApJ...312..860K, Vos2011, 2013ApJ...774...72K}.

\section{Summary \& Conclusions}\label{sec:summary}
In this work we set off from a uniform data set sampling 54 strong and commonly detected DIBs in 25 sight lines representing various ISM environments. We investigate their pair-wise correlations, and provide further analysis using data science tools like clustering and MDS. The major conclusions we have reached are as follows:

\begin{enumerate}
    \item Using normalized equivalent widths $W$(DIBs)/$E_\textrm{B-V}$, we confirm the common presence of anti-correlated DIB pairs, most notably between the C$_2$ DIBs and the $\sigma$-type DIBs like $\lambda\lambda$5780 and 6284. We find several DIB pairs with $r_\textrm{norm} \geq 0.9$, but the most negative $r_\textrm{norm}$ values are $\sim -0.7$, far from a perfect anti-correlation. We do not see convincing evidence that any of the anti-correlated DIB pairs is from successive ionization states of a single carrier.
    \item We use multi-dimensional scaling (MDS) to visualize the 1 - $r_\textrm{norm}$ dissimilarities as distances between data points. At least two projection axes are required to properly explain the dissimilarities among DIBs. The first factor is associated with radiation. The second factor is not fully understood and might be related to the interaction of DIB carriers with other particles, species, or dust grains.
    \item Hierarchical agglomerative clustering and $k$-means clustering reproduce previous divisions of DIB families, including the $\sigma$-, $\zeta$-, and C$_2$ DIBs. However, the MDS analysis shows a rather continuous sequence especially among the non-C$_2$ DIBs, and the $\sigma$- and $\zeta$-type DIBs appear to be the two extreme ends of this continuous sequence. The continuous sequence suggests the DIB carriers differ by a property that is continuously variable, such as ionization potential and maybe molecular size. There is a gap between the non-C$_2$ DIBs and the C$_2$ DIBs which suggests more fundamental differences between the carriers of the two groups of DIBs.
\end{enumerate}

\section*{Acknowledgements}

HF, MS and JC acknowledge support from an NSERC Discovery Grant, a Western SERB Accelerator Award and the USRI program. HF would like to thank Prof. Gang Zhao from NAOC and Prof. Donald G. York from the UofChicago for their supervision and kind support during his PhD study. The hierarchical agglomerative clustering (HAC), $k$-means clustering, and multi-dimensional scaling (MDS) analyses in this work are based on the scikit-learn package \citep{scikit-learn}. This research has made use of NASA's Astrophysics Data System Bibliographic Services and the SIMBAD database operated at CDS, Strasbourg, France \citep{2000A&AS..143....9W}.

\section*{Data Availability}
DIB measurements used in this work are taken from \cite{APOcat}. These data are open-access at \url{https://cdsarc.unistra.fr/viz-bin/cat/J/ApJ/878/151}. The code used to generate the figures are available at \url{https://github.com/HaoyuFan-DIB/FamilyAndClusterOfDIBs}.
  



\bibliographystyle{mnras}
\bibliography{Main} 

\begin{thebibliography}{}
\makeatletter
\relax
\def\mn@urlcharsother{\let\do\@makeother \do\$\do\&\do\#\do\^\do\_\do\%\do\~}
\def\mn@doi{\begingroup\mn@urlcharsother \@ifnextchar [ {\mn@doi@}
  {\mn@doi@[]}}
\def\mn@doi@[#1]#2{\def\@tempa{#1}\ifx\@tempa\@empty \href
  {http://dx.doi.org/#2} {doi:#2}\else \href {http://dx.doi.org/#2} {#1}\fi
  \endgroup}
\def\mn@eprint#1#2{\mn@eprint@#1:#2::\@nil}
\def\mn@eprint@arXiv#1{\href {http://arxiv.org/abs/#1} {{\tt arXiv:#1}}}
\def\mn@eprint@dblp#1{\href {http://dblp.uni-trier.de/rec/bibtex/#1.xml}
  {dblp:#1}}
\def\mn@eprint@#1:#2:#3:#4\@nil{\def\@tempa {#1}\def\@tempb {#2}\def\@tempc
  {#3}\ifx \@tempc \@empty \let \@tempc \@tempb \let \@tempb \@tempa \fi \ifx
  \@tempb \@empty \def\@tempb {arXiv}\fi \@ifundefined
  {mn@eprint@\@tempb}{\@tempb:\@tempc}{\expandafter \expandafter \csname
  mn@eprint@\@tempb\endcsname \expandafter{\@tempc}}}

\bibitem[\protect\citeauthoryear{{Adamson}, {Whittet}  \& {Duley}}{{Adamson}
  et~al.}{1991}]{1991MNRAS.252..234A}
{Adamson} A.~J.,  {Whittet} D.~C.~B.,   {Duley} W.~W.,  1991, \mn@doi [\mnras]
  {10.1093/mnras/252.2.234}, \href
  {https://ui-adsabs-harvard-edu.proxy1.lib.uwo.ca/abs/1991MNRAS.252..234A}
  {252, 234}

\bibitem[\protect\citeauthoryear{{Allain}, {Leach}  \& {Sedlmayr}}{{Allain}
  et~al.}{1996}]{1996A&A...305..616A}
{Allain} T.,  {Leach} S.,   {Sedlmayr} E.,  1996, \aap, \href
  {https://ui-adsabs-harvard-edu.proxy1.lib.uwo.ca/abs/1996A&A...305..616A}
  {305, 616}

\bibitem[\protect\citeauthoryear{{Andrews}, {Peeters}, {Tielens}  \&
  {Okada}}{{Andrews} et~al.}{2018}]{2018A&A...619A.170A}
{Andrews} H.,  {Peeters} E.,  {Tielens} A.~G.~G.~M.,   {Okada} Y.,  2018,
  \mn@doi [\aap] {10.1051/0004-6361/201832808}, \href
  {https://ui-adsabs-harvard-edu.proxy1.lib.uwo.ca/abs/2018A&A...619A.170A}
  {619, A170}

\bibitem[\protect\citeauthoryear{{Bailey}, {van Loon}, {Sarre}  \&
  {Beckman}}{{Bailey} et~al.}{2015}]{2015MNRAS.454.4013B}
{Bailey} M.,  {van Loon} J.~T.,  {Sarre} P.~J.,   {Beckman} J.~E.,  2015,
  \mn@doi [\mnras] {10.1093/mnras/stv2178}, \href
  {https://ui-adsabs-harvard-edu.proxy1.lib.uwo.ca/abs/2015MNRAS.454.4013B}
  {454, 4013}

\bibitem[\protect\citeauthoryear{{Bailey}, {van Loon}, {Farhang}, {Javadi},
  {Khosroshahi}, {Sarre}  \& {Smith}}{{Bailey}
  et~al.}{2016}]{2016A&A...585A..12B}
{Bailey} M.,  {van Loon} J.~T.,  {Farhang} A.,  {Javadi} A.,  {Khosroshahi}
  H.~G.,  {Sarre} P.~J.,   {Smith} K.~T.,  2016, \mn@doi [\aap]
  {10.1051/0004-6361/201526656}, \href
  {https://ui-adsabs-harvard-edu.proxy1.lib.uwo.ca/abs/2016A&A...585A..12B}
  {585, A12}

\bibitem[\protect\citeauthoryear{{Baron}, {Poznanski}, {Watson}, {Yao}, {Cox}
  \& {Prochaska}}{{Baron} et~al.}{2015}]{2015MNRAS.451..332B}
{Baron} D.,  {Poznanski} D.,  {Watson} D.,  {Yao} Y.,  {Cox} N. L.~J.,
  {Prochaska} J.~X.,  2015, \mn@doi [\mnras] {10.1093/mnras/stv977}, \href
  {https://ui-adsabs-harvard-edu.proxy1.lib.uwo.ca/abs/2015MNRAS.451..332B}
  {451, 332}

\bibitem[\protect\citeauthoryear{Borg \& Groenen}{Borg \& Groenen}{2005}]{MDS2}
Borg I.,  Groenen P.,  2005, Modern Multidimensional Scaling: Theory and
  Applications.
Springer Series in Statistics, Springer New York, \url
  {https://books.google.ca/books?id=duTODldZzRcC}

\bibitem[\protect\citeauthoryear{{Cami}, {Sonnentrucker}, {Ehrenfreund}  \&
  {Foing}}{{Cami} et~al.}{1997}]{Cami:DIBcorrelations}
{Cami} J.,  {Sonnentrucker} P.,  {Ehrenfreund} P.,   {Foing} B.~H.,  1997,
  \aap, 326, 822

\bibitem[\protect\citeauthoryear{{Cami}, {Salama}, {Jim{\'e}nez-Vicente},
  {Galazutdinov}  \& {Kre{\l}owski}}{{Cami} et~al.}{2004}]{2004ApJ...611L.113C}
{Cami} J.,  {Salama} F.,  {Jim{\'e}nez-Vicente} J.,  {Galazutdinov} G.~A.,
  {Kre{\l}owski} J.,  2004, \mn@doi [\apjl] {10.1086/423991}, \href
  {https://ui-adsabs-harvard-edu.proxy1.lib.uwo.ca/abs/2004ApJ...611L.113C}
  {611, L113}

\bibitem[\protect\citeauthoryear{{Campbell}, {Holz}, {Gerlich}  \&
  {Maier}}{{Campbell} et~al.}{2015}]{2015Natur.523..322C}
{Campbell} E.~K.,  {Holz} M.,  {Gerlich} D.,   {Maier} J.~P.,  2015, \mn@doi
  [\nat] {10.1038/nature14566}, \href
  {https://ui-adsabs-harvard-edu.proxy1.lib.uwo.ca/abs/2015Natur.523..322C}
  {523, 322}

\bibitem[\protect\citeauthoryear{{Cardelli}}{{Cardelli}}{1994}]{1994Sci...265..209C}
{Cardelli} J.~A.,  1994, \mn@doi [Science] {10.1126/science.265.5169.209},
  \href
  {https://ui-adsabs-harvard-edu.proxy1.lib.uwo.ca/abs/1994Sci...265..209C}
  {265, 209}

\bibitem[\protect\citeauthoryear{{Cordiner} et~al.,}{{Cordiner}
  et~al.}{2017}]{2017ApJ...843L...2C}
{Cordiner} M.~A.,  et~al., 2017, \mn@doi [\apjl] {10.3847/2041-8213/aa78f7},
  \href {http://adsabs.harvard.edu/abs/2017ApJ...843L...2C} {843, L2}

\bibitem[\protect\citeauthoryear{{Cordiner} et~al.,}{{Cordiner}
  et~al.}{2019}]{2019ApJ...875L..28C}
{Cordiner} M.~A.,  et~al., 2019, \mn@doi [\apjl] {10.3847/2041-8213/ab14e5},
  \href {https://ui.adsabs.harvard.edu/abs/2019ApJ...875L..28C} {875, L28}

\bibitem[\protect\citeauthoryear{{Cox} \& {Spaans}}{{Cox} \&
  {Spaans}}{2006}]{2006A&A...451..973C}
{Cox} N.~L.~J.,  {Spaans} M.,  2006, \mn@doi [\aap]
  {10.1051/0004-6361:20054484}, \href
  {https://ui-adsabs-harvard-edu.proxy1.lib.uwo.ca/abs/2006A&A...451..973C}
  {451, 973}

\bibitem[\protect\citeauthoryear{{Cox}, {Cordiner}, {Cami}, {Foing}, {Sarre},
  {Kaper}  \& {Ehrenfreund}}{{Cox} et~al.}{2006}]{Cox2006}
{Cox} N.~L.~J.,  {Cordiner} M.~A.,  {Cami} J.,  {Foing} B.~H.,  {Sarre} P.~J.,
  {Kaper} L.,   {Ehrenfreund} P.,  2006, \mn@doi [\aap]
  {10.1051/0004-6361:20053367}, \href
  {https://ui-adsabs-harvard-edu.proxy1.lib.uwo.ca/abs/2006A&A...447..991C}
  {447, 991}

\bibitem[\protect\citeauthoryear{{Cox}, {Ehrenfreund}, {Foing}, {D'Hendecourt},
  {Salama}  \& {Sarre}}{{Cox} et~al.}{2011}]{Cox2011}
{Cox} N.~L.~J.,  {Ehrenfreund} P.,  {Foing} B.~H.,  {D'Hendecourt} L.,
  {Salama} F.,   {Sarre} P.~J.,  2011, \mn@doi [\aap]
  {10.1051/0004-6361/201016365}, \href
  {https://ui-adsabs-harvard-edu.proxy1.lib.uwo.ca/abs/2011A&A...531A..25C}
  {531, A25}

\bibitem[\protect\citeauthoryear{{Cox}, {Cami}, {Kaper}, {Ehrenfreund},
  {Foing}, {Ochsendorf}, {van Hooff}  \& {Salama}}{{Cox}
  et~al.}{2014}]{2014A&A...569A.117C}
{Cox} N.~L.~J.,  {Cami} J.,  {Kaper} L.,  {Ehrenfreund} P.,  {Foing} B.~H.,
  {Ochsendorf} B.~B.,  {van Hooff} S.~H.~M.,   {Salama} F.,  2014, \mn@doi
  [\aap] {10.1051/0004-6361/201323061}, \href
  {https://ui-adsabs-harvard-edu.proxy1.lib.uwo.ca/abs/2014A&A...569A.117C}
  {569, A117}

\bibitem[\protect\citeauthoryear{{Cox} et~al.,}{{Cox}
  et~al.}{2017}]{2017A&A...606A..76C}
{Cox} N. L.~J.,  et~al., 2017, \mn@doi [\aap] {10.1051/0004-6361/201730912},
  \href {https://ui.adsabs.harvard.edu/abs/2017A&A...606A..76C} {606, A76}

\bibitem[\protect\citeauthoryear{{Dahlstrom} et~al.,}{{Dahlstrom}
  et~al.}{2013}]{2013ApJ...773...41D}
{Dahlstrom} J.,  et~al., 2013, \mn@doi [\apj] {10.1088/0004-637X/773/1/41},
  \href
  {https://ui-adsabs-harvard-edu.proxy1.lib.uwo.ca/abs/2013ApJ...773...41D}
  {773, 41}

\bibitem[\protect\citeauthoryear{{De Cia}, {Jenkins}, {Fox}, {Ledoux},
  {Ramburth-Hurt}, {Konstantopoulou}, {Petitjean}  \& {Krogager}}{{De Cia}
  et~al.}{2021}]{2021Natur.597..206D}
{De Cia} A.,  {Jenkins} E.~B.,  {Fox} A.~J.,  {Ledoux} C.,  {Ramburth-Hurt} T.,
   {Konstantopoulou} C.,  {Petitjean} P.,   {Krogager} J.-K.,  2021, \mn@doi
  [\nat] {10.1038/s41586-021-03780-0}, \href
  {https://ui-adsabs-harvard-edu.proxy1.lib.uwo.ca/abs/2021Natur.597..206D}
  {597, 206}

\bibitem[\protect\citeauthoryear{{Ehrenfreund} \& {Foing}}{{Ehrenfreund} \&
  {Foing}}{1996}]{1996A&A...307L..25E}
{Ehrenfreund} P.,  {Foing} B.~H.,  1996, \aap, \href
  {https://ui-adsabs-harvard-edu.proxy1.lib.uwo.ca/abs/1996A&A...307L..25E}
  {307, L25}

\bibitem[\protect\citeauthoryear{{Ehrenfreund}, {Cami}, {Dartois}  \&
  {Foing}}{{Ehrenfreund} et~al.}{1997}]{1997A&A...318L..28E}
{Ehrenfreund} P.,  {Cami} J.,  {Dartois} E.,   {Foing} B.~H.,  1997, \aap,
  \href
  {https://ui-adsabs-harvard-edu.proxy1.lib.uwo.ca/abs/1997A&A...318L..28E}
  {318, L28}

\bibitem[\protect\citeauthoryear{{Elyajouri} et~al.,}{{Elyajouri}
  et~al.}{2018}]{EDIBLESIII_C2}
{Elyajouri} M.,  et~al., 2018, \mn@doi [\aap] {10.1051/0004-6361/201833105},
  \href
  {https://ui-adsabs-harvard-edu.proxy1.lib.uwo.ca/abs/2018A&A...616A.143E}
  {616, A143}

\bibitem[\protect\citeauthoryear{{Ensor}, {Cami}, {Bhatt}  \& {Soddu}}{{Ensor}
  et~al.}{2017}]{PCA_Cami}
{Ensor} T.,  {Cami} J.,  {Bhatt} N.~H.,   {Soddu} A.,  2017, \mn@doi [\apj]
  {10.3847/1538-4357/aa5b84}, \href
  {https://ui-adsabs-harvard-edu.proxy1.lib.uwo.ca/abs/2017ApJ...836..162E}
  {836, 162}

\bibitem[\protect\citeauthoryear{{Fan} et~al.,}{{Fan} et~al.}{2017}]{behavior}
{Fan} H.,  et~al., 2017, \mn@doi [\apj] {10.3847/1538-4357/aa9480}, \href
  {https://ui-adsabs-harvard-edu.proxy1.lib.uwo.ca/abs/2017ApJ...850..194F}
  {850, 194}

\bibitem[\protect\citeauthoryear{{Fan} et~al.,}{{Fan} et~al.}{2019}]{APOcat}
{Fan} H.,  et~al., 2019, \mn@doi [\apj] {10.3847/1538-4357/ab1b74}, \href
  {https://ui-adsabs-harvard-edu.proxy1.lib.uwo.ca/abs/2019ApJ...878..151F}
  {878, 151}

\bibitem[\protect\citeauthoryear{{Farhang}, {van Loon}, {Khosroshahi}, {Javadi}
   \& {Bailey}}{{Farhang} et~al.}{2019}]{2019NatAs...3..922F}
{Farhang} A.,  {van Loon} J.~T.,  {Khosroshahi} H.~G.,  {Javadi} A.,   {Bailey}
  M.,  2019, \mn@doi [Nature Astronomy] {10.1038/s41550-019-0814-z}, \href
  {https://ui-adsabs-harvard-edu.proxy1.lib.uwo.ca/abs/2019NatAs...3..922F} {3,
  922}

\bibitem[\protect\citeauthoryear{{Foing} \& {Ehrenfreund}}{{Foing} \&
  {Ehrenfreund}}{1994}]{1994Natur.369..296F}
{Foing} B.~H.,  {Ehrenfreund} P.,  1994, \mn@doi [\nat] {10.1038/369296a0},
  \href
  {https://ui-adsabs-harvard-edu.proxy1.lib.uwo.ca/abs/1994Natur.369..296F}
  {369, 296}

\bibitem[\protect\citeauthoryear{{Foing} \& {Ehrenfreund}}{{Foing} \&
  {Ehrenfreund}}{1997}]{1997A&A...317L..59F}
{Foing} B.~H.,  {Ehrenfreund} P.,  1997, \aap, \href
  {https://ui-adsabs-harvard-edu.proxy1.lib.uwo.ca/abs/1997A&A...317L..59F}
  {317, L59}

\bibitem[\protect\citeauthoryear{{Friedman} et~al.,}{{Friedman}
  et~al.}{2011}]{Friedman2011}
{Friedman} S.~D.,  et~al., 2011, \mn@doi [\apj] {10.1088/0004-637X/727/1/33},
  \href
  {https://ui-adsabs-harvard-edu.proxy1.lib.uwo.ca/abs/2011ApJ...727...33F}
  {727, 33}

\bibitem[\protect\citeauthoryear{{Galazutdinov}, {Moutou}, {Musaev}  \&
  {Kre{\l}owski}}{{Galazutdinov} et~al.}{2002}]{2002A&A...384..215G}
{Galazutdinov} G.,  {Moutou} C.,  {Musaev} F.,   {Kre{\l}owski} J.,  2002,
  \mn@doi [\aap] {10.1051/0004-6361:20020003}, \href
  {https://ui-adsabs-harvard-edu.proxy1.lib.uwo.ca/abs/2002A&A...384..215G}
  {384, 215}

\bibitem[\protect\citeauthoryear{{Galazutdinov}, {Musaev}, {Bondar}  \&
  {Kre{\l}owski}}{{Galazutdinov} et~al.}{2003}]{2003MNRAS.345..365G}
{Galazutdinov} G.~A.,  {Musaev} F.~A.,  {Bondar} A.~V.,   {Kre{\l}owski} J.,
  2003, \mn@doi [\mnras] {10.1046/j.1365-8711.2003.06966.x}, \href
  {https://ui-adsabs-harvard-edu.proxy1.lib.uwo.ca/abs/2003MNRAS.345..365G}
  {345, 365}

\bibitem[\protect\citeauthoryear{{Galazutdinov}, {Gnaci{\'n}ski}, {Han}, {Lee},
  {Kim}  \& {Kre{\l}owski}}{{Galazutdinov} et~al.}{2006}]{2006A&A...447..589G}
{Galazutdinov} G.~A.,  {Gnaci{\'n}ski} P.,  {Han} I.,  {Lee} B.-C.,  {Kim}
  K.-M.,   {Kre{\l}owski} J.,  2006, \mn@doi [\aap]
  {10.1051/0004-6361:20053410}, \href
  {https://ui-adsabs-harvard-edu.proxy1.lib.uwo.ca/abs/2006A&A...447..589G}
  {447, 589}

\bibitem[\protect\citeauthoryear{{Galazutdinov}, {Valyavin}, {Ikhsanov}  \&
  {Kre{\l}owski}}{{Galazutdinov} et~al.}{2021}]{2021AJ....161..127G}
{Galazutdinov} G.~A.,  {Valyavin} G.,  {Ikhsanov} N.~R.,   {Kre{\l}owski} J.,
  2021, \mn@doi [\aj] {10.3847/1538-3881/abd4e5}, \href
  {https://ui-adsabs-harvard-edu.proxy1.lib.uwo.ca/abs/2021AJ....161..127G}
  {161, 127}

\bibitem[\protect\citeauthoryear{{Gonz{\'a}lez Hern{\'a}ndez},
  {Iglesias-Groth}, {Rebolo}, {Garc{\'\i}a-Hern{\'a}ndez}, {Manchado}  \&
  {Lambert}}{{Gonz{\'a}lez Hern{\'a}ndez} et~al.}{2009}]{2009ApJ...706..866G}
{Gonz{\'a}lez Hern{\'a}ndez} J.~I.,  {Iglesias-Groth} S.,  {Rebolo} R.,
  {Garc{\'\i}a-Hern{\'a}ndez} D.~A.,  {Manchado} A.,   {Lambert} D.~L.,  2009,
  \mn@doi [\apj] {10.1088/0004-637X/706/1/866}, \href
  {https://ui-adsabs-harvard-edu.proxy1.lib.uwo.ca/abs/2009ApJ...706..866G}
  {706, 866}

\bibitem[\protect\citeauthoryear{{Hamano} et~al.,}{{Hamano}
  et~al.}{2015}]{2015ApJ...800..137H}
{Hamano} S.,  et~al., 2015, \mn@doi [\apj] {10.1088/0004-637X/800/2/137}, \href
  {https://ui-adsabs-harvard-edu.proxy1.lib.uwo.ca/abs/2015ApJ...800..137H}
  {800, 137}

\bibitem[\protect\citeauthoryear{{Hamano} et~al.,}{{Hamano}
  et~al.}{2016}]{2016ApJ...821...42H}
{Hamano} S.,  et~al., 2016, \mn@doi [\apj] {10.3847/0004-637X/821/1/42}, \href
  {https://ui-adsabs-harvard-edu.proxy1.lib.uwo.ca/abs/2016ApJ...821...42H}
  {821, 42}

\bibitem[\protect\citeauthoryear{{Heger}}{{Heger}}{1922}]{1922LicOB..10..146H}
{Heger} M.~L.,  1922, Lick Observatory Bulletin, \href
  {https://ui-adsabs-harvard-edu.proxy1.lib.uwo.ca/abs/1922LicOB..10..146H}
  {10, 146}

\bibitem[\protect\citeauthoryear{{Herbig}}{{Herbig}}{1993}]{Herbig1993}
{Herbig} G.~H.,  1993, \mn@doi [\apj] {10.1086/172500}, \href
  {https://ui-adsabs-harvard-edu.proxy1.lib.uwo.ca/abs/1993ApJ...407..142H}
  {407, 142}

\bibitem[\protect\citeauthoryear{{Herbig}}{{Herbig}}{1995}]{1995ARA&A..33...19H}
{Herbig} G.~H.,  1995, \mn@doi [\araa] {10.1146/annurev.aa.33.090195.000315},
  \href
  {https://ui-adsabs-harvard-edu.proxy1.lib.uwo.ca/abs/1995ARA&A..33...19H}
  {33, 19}

\bibitem[\protect\citeauthoryear{{Hobbs} et~al.,}{{Hobbs}
  et~al.}{2008}]{2008ApJ...680.1256H}
{Hobbs} L.~M.,  et~al., 2008, \mn@doi [\apj] {10.1086/587930}, \href
  {https://ui-adsabs-harvard-edu.proxy1.lib.uwo.ca/abs/2008ApJ...680.1256H}
  {680, 1256}

\bibitem[\protect\citeauthoryear{{Hobbs} et~al.,}{{Hobbs}
  et~al.}{2009}]{2009ApJ...705...32H}
{Hobbs} L.~M.,  et~al., 2009, \mn@doi [\apj] {10.1088/0004-637X/705/1/32},
  \href
  {https://ui-adsabs-harvard-edu.proxy1.lib.uwo.ca/abs/2009ApJ...705...32H}
  {705, 32}

\bibitem[\protect\citeauthoryear{{Jenniskens}, {Ehrenfreund}  \&
  {Foing}}{{Jenniskens} et~al.}{1994}]{1994A&A...281..517J}
{Jenniskens} P.,  {Ehrenfreund} P.,   {Foing} B.,  1994, \aap, \href
  {https://ui.adsabs.harvard.edu/abs/1994A&A...281..517J} {281, 517}

\bibitem[\protect\citeauthoryear{{Jensen} \& {Snow}}{{Jensen} \&
  {Snow}}{2007a}]{2007ApJ...669..378J}
{Jensen} A.~G.,  {Snow} T.~P.,  2007a, \mn@doi [\apj] {10.1086/521638}, \href
  {https://ui-adsabs-harvard-edu.proxy1.lib.uwo.ca/abs/2007ApJ...669..378J}
  {669, 378}

\bibitem[\protect\citeauthoryear{{Jensen} \& {Snow}}{{Jensen} \&
  {Snow}}{2007b}]{2007ApJ...669..401J}
{Jensen} A.~G.,  {Snow} T.~P.,  2007b, \mn@doi [\apj] {10.1086/521420}, \href
  {https://ui-adsabs-harvard-edu.proxy1.lib.uwo.ca/abs/2007ApJ...669..401J}
  {669, 401}

\bibitem[\protect\citeauthoryear{{Joblin}, {Maillard}, {D'Hendecourt}  \&
  {L{\'e}ger}}{{Joblin} et~al.}{1990}]{1990Natur.346..729J}
{Joblin} C.,  {Maillard} J.~P.,  {D'Hendecourt} L.,   {L{\'e}ger} A.,  1990,
  \mn@doi [\nat] {10.1038/346729a0}, \href
  {https://ui-adsabs-harvard-edu.proxy1.lib.uwo.ca/abs/1990Natur.346..729J}
  {346, 729}

\bibitem[\protect\citeauthoryear{{Josafatsson} \& {Snow}}{{Josafatsson} \&
  {Snow}}{1987}]{1987ApJ...319..436J}
{Josafatsson} K.,  {Snow} T.~P.,  1987, \mn@doi [\apj] {10.1086/165468}, \href
  {https://ui-adsabs-harvard-edu.proxy1.lib.uwo.ca/abs/1987ApJ...319..436J}
  {319, 436}

\bibitem[\protect\citeauthoryear{{Kerr}, {Hibbins}, {Fossey}, {Miles}  \&
  {Sarre}}{{Kerr} et~al.}{1998}]{1998ApJ...495..941K}
{Kerr} T.~H.,  {Hibbins} R.~E.,  {Fossey} S.~J.,  {Miles} J.~R.,   {Sarre}
  P.~J.,  1998, \mn@doi [\apj] {10.1086/305339}, \href
  {https://ui-adsabs-harvard-edu.proxy1.lib.uwo.ca/abs/1998ApJ...495..941K}
  {495, 941}

\bibitem[\protect\citeauthoryear{{Kos} \& {Zwitter}}{{Kos} \&
  {Zwitter}}{2013}]{2013ApJ...774...72K}
{Kos} J.,  {Zwitter} T.,  2013, \mn@doi [\apj] {10.1088/0004-637X/774/1/72},
  \href
  {https://ui-adsabs-harvard-edu.proxy1.lib.uwo.ca/abs/2013ApJ...774...72K}
  {774, 72}

\bibitem[\protect\citeauthoryear{{Kre{\l}owski}}{{Kre{\l}owski}}{2018}]{2018PASP..130g1001K}
{Kre{\l}owski} J.,  2018, \mn@doi [\pasp] {10.1088/1538-3873/aabd69}, \href
  {https://ui-adsabs-harvard-edu.proxy1.lib.uwo.ca/abs/2018PASP..130g1001K}
  {130, 071001}

\bibitem[\protect\citeauthoryear{{Krelowski} \& {Walker}}{{Krelowski} \&
  {Walker}}{1987}]{1987ApJ...312..860K}
{Krelowski} J.,  {Walker} G.~A.~H.,  1987, \mn@doi [\apj] {10.1086/164932},
  \href
  {https://ui-adsabs-harvard-edu.proxy1.lib.uwo.ca/abs/1987ApJ...312..860K}
  {312, 860}

\bibitem[\protect\citeauthoryear{{Lai}, {Witt}, {Alvarez}  \& {Cami}}{{Lai}
  et~al.}{2020}]{2020MNRAS.492.5853L}
{Lai} T. S.~Y.,  {Witt} A.~N.,  {Alvarez} C.,   {Cami} J.,  2020, \mn@doi
  [\mnras] {10.1093/mnras/staa223}, \href
  {https://ui-adsabs-harvard-edu.proxy1.lib.uwo.ca/abs/2020MNRAS.492.5853L}
  {492, 5853}

\bibitem[\protect\citeauthoryear{{Lallement} et~al.,}{{Lallement}
  et~al.}{2018}]{Lallement:EDIBLES2}
{Lallement} R.,  et~al., 2018, \mn@doi [\aap] {10.1051/0004-6361/201832647},
  \href {http://adsabs.harvard.edu/abs/2018A%26A...614A..28L} {614, A28}

\bibitem[\protect\citeauthoryear{{Lan}, {M{\'e}nard}  \& {Zhu}}{{Lan}
  et~al.}{2015}]{2015MNRAS.452.3629L}
{Lan} T.-W.,  {M{\'e}nard} B.,   {Zhu} G.,  2015, \mn@doi [\mnras]
  {10.1093/mnras/stv1519}, \href
  {https://ui-adsabs-harvard-edu.proxy1.lib.uwo.ca/abs/2015MNRAS.452.3629L}
  {452, 3629}

\bibitem[\protect\citeauthoryear{Linnartz, Cami, Cordiner, Cox, Ehrenfreund,
  Foing, Gatchell  \& Scheier}{Linnartz et~al.}{2020}]{Harold:C60plusreview}
Linnartz H.,  Cami J.,  Cordiner M.,  Cox N.,  Ehrenfreund P.,  Foing B.,
  Gatchell M.,   Scheier P.,  2020, \mn@doi [Journal of Molecular Spectroscopy]
  {https://doi.org/10.1016/j.jms.2019.111243}, \href
  {http://www.sciencedirect.com/science/article/pii/S0022285219303182} {367,
  111243}

\bibitem[\protect\citeauthoryear{{McCall}, {Thorburn}, {Hobbs}, {Oka}  \&
  {York}}{{McCall} et~al.}{2001}]{2001ApJ...559L..49M}
{McCall} B.~J.,  {Thorburn} J.,  {Hobbs} L.~M.,  {Oka} T.,   {York} D.~G.,
  2001, \mn@doi [\apjl] {10.1086/323669}, \href
  {https://ui-adsabs-harvard-edu.proxy1.lib.uwo.ca/abs/2001ApJ...559L..49M}
  {559, L49}

\bibitem[\protect\citeauthoryear{{McCall} et~al.,}{{McCall}
  et~al.}{2010}]{6196-6613}
{McCall} B.~J.,  et~al., 2010, \mn@doi [\apj] {10.1088/0004-637X/708/2/1628},
  \href
  {https://ui-adsabs-harvard-edu.proxy1.lib.uwo.ca/abs/2010ApJ...708.1628M}
  {708, 1628}

\bibitem[\protect\citeauthoryear{{McInnes}, {Healy}  \& {Melville}}{{McInnes}
  et~al.}{2018}]{2018arXivUMAP}
{McInnes} L.,  {Healy} J.,   {Melville} J.,  2018, preprint (\mn@eprint {arXiv}
  {1802.03426})

\bibitem[\protect\citeauthoryear{Mead}{Mead}{1992}]{MDS1}
Mead A.,  1992, Journal of the Royal Statistical Society. Series D (The
  Statistician), 41, 27

\bibitem[\protect\citeauthoryear{{Meyer} \& {Ulrich}}{{Meyer} \&
  {Ulrich}}{1984}]{1984ApJ...283...98M}
{Meyer} D.~M.,  {Ulrich} R.~K.,  1984, \mn@doi [\apj] {10.1086/162278}, \href
  {https://ui-adsabs-harvard-edu.proxy1.lib.uwo.ca/abs/1984ApJ...283...98M}
  {283, 98}

\bibitem[\protect\citeauthoryear{{Moutou}, {Kre{\l}owski}, {D'Hendecourt}  \&
  {Jamroszczak}}{{Moutou} et~al.}{1999}]{1999A&A...351..680M}
{Moutou} C.,  {Kre{\l}owski} J.,  {D'Hendecourt} L.,   {Jamroszczak} J.,  1999,
  \aap, \href
  {https://ui-adsabs-harvard-edu.proxy1.lib.uwo.ca/abs/1999A&A...351..680M}
  {351, 680}

\bibitem[\protect\citeauthoryear{{Omont} \& {Bettinger}}{{Omont} \&
  {Bettinger}}{2020}]{2020A&A...637A..74O}
{Omont} A.,  {Bettinger} H.~F.,  2020, \mn@doi [\aap]
  {10.1051/0004-6361/201937071}, \href
  {https://ui-adsabs-harvard-edu.proxy1.lib.uwo.ca/abs/2020A&A...637A..74O}
  {637, A74}

\bibitem[\protect\citeauthoryear{{Omont} \& {Bettinger}}{{Omont} \&
  {Bettinger}}{2021}]{2021A&A...650A.193O}
{Omont} A.,  {Bettinger} H.~F.,  2021, \mn@doi [\aap]
  {10.1051/0004-6361/202140675}, \href
  {https://ui-adsabs-harvard-edu.proxy1.lib.uwo.ca/abs/2021A&A...650A.193O}
  {650, A193}

\bibitem[\protect\citeauthoryear{Pedregosa et~al.,}{Pedregosa
  et~al.}{2011}]{scikit-learn}
Pedregosa F.,  et~al., 2011, Journal of Machine Learning Research, 12, 2825

\bibitem[\protect\citeauthoryear{{Ram{\'\i}rez-Tannus}, {Cox}, {Kaper}  \& {de
  Koter}}{{Ram{\'\i}rez-Tannus} et~al.}{2018}]{2018A&A...620A..52R}
{Ram{\'\i}rez-Tannus} M.~C.,  {Cox} N.~L.~J.,  {Kaper} L.,   {de Koter} A.,
  2018, \mn@doi [\aap] {10.1051/0004-6361/201833340}, \href
  {https://ui.adsabs.harvard.edu/abs/2018A&A...620A..52R} {620, A52}

\bibitem[\protect\citeauthoryear{{Rawlings}, {Juvela}, {Lehtinen}, {Mattila}
  \& {Lemke}}{{Rawlings} et~al.}{2013}]{2013MNRAS.428.2617R}
{Rawlings} M.~G.,  {Juvela} M.,  {Lehtinen} K.,  {Mattila} K.,   {Lemke} D.,
  2013, \mn@doi [\mnras] {10.1093/mnras/sts233}, \href
  {https://ui-adsabs-harvard-edu.proxy1.lib.uwo.ca/abs/2013MNRAS.428.2617R}
  {428, 2617}

\bibitem[\protect\citeauthoryear{{Sarre}, {Miles}, {Kerr}, {Hibbins}, {Fossey}
  \& {Somerville}}{{Sarre} et~al.}{1995}]{1995MNRAS.277L..41S}
{Sarre} P.~J.,  {Miles} J.~R.,  {Kerr} T.~H.,  {Hibbins} R.~E.,  {Fossey}
  S.~J.,   {Somerville} W.~B.,  1995, \mn@doi [\mnras]
  {10.1093/mnras/277.1.L41}, \href
  {https://ui-adsabs-harvard-edu.proxy1.lib.uwo.ca/abs/1995MNRAS.277L..41S}
  {277, L41}

\bibitem[\protect\citeauthoryear{{Savage}, {Bohlin}, {Drake}  \&
  {Budich}}{{Savage} et~al.}{1977}]{1977ApJ...216..291S}
{Savage} B.~D.,  {Bohlin} R.~C.,  {Drake} J.~F.,   {Budich} W.,  1977, \mn@doi
  [\apj] {10.1086/155471}, \href
  {https://ui-adsabs-harvard-edu.proxy1.lib.uwo.ca/abs/1977ApJ...216..291S}
  {216, 291}

\bibitem[\protect\citeauthoryear{{Smith}, {Harriott}, {Majaess}, {Massa}  \&
  {Matta}}{{Smith} et~al.}{2021}]{2021MNRAS.507.5236S}
{Smith} F.~M.,  {Harriott} T.~A.,  {Majaess} D.,  {Massa} L.,   {Matta} C.~F.,
  2021, \mn@doi [\mnras] {10.1093/mnras/stab2444}, \href
  {https://ui-adsabs-harvard-edu.proxy1.lib.uwo.ca/abs/2021MNRAS.507.5236S}
  {507, 5236}

\bibitem[\protect\citeauthoryear{{Snow} \& {Cohen}}{{Snow} \&
  {Cohen}}{1974}]{1974ApJ...194..313S}
{Snow} T.~P. J.,  {Cohen} J.~G.,  1974, \mn@doi [\apj] {10.1086/153247}, \href
  {https://ui.adsabs.harvard.edu/abs/1974ApJ...194..313S} {194, 313}

\bibitem[\protect\citeauthoryear{{Sonnentrucker}}{{Sonnentrucker}}{2014}]{2014IAUS..297...13S}
{Sonnentrucker} P.,  2014, in {Cami} J.,  {Cox} N. L.~J.,  eds,  IAU Symposium
  Vol. 297, The Diffuse Interstellar Bands. pp 13--22,
  \mn@doi{10.1017/S1743921313015524}

\bibitem[\protect\citeauthoryear{{Sonnentrucker}, {Cami}, {Ehrenfreund}  \&
  {Foing}}{{Sonnentrucker} et~al.}{1997}]{1997A&A...327.1215S}
{Sonnentrucker} P.,  {Cami} J.,  {Ehrenfreund} P.,   {Foing} B.~H.,  1997,
  \aap, \href
  {https://ui-adsabs-harvard-edu.proxy1.lib.uwo.ca/abs/1997A&A...327.1215S}
  {327, 1215}

\bibitem[\protect\citeauthoryear{{Sonnentrucker}, {York}, {Hobbs}, {Welty},
  {Friedman}, {Dahlstrom}, {Snow}  \& {York}}{{Sonnentrucker}
  et~al.}{2018}]{broadDIB}
{Sonnentrucker} P.,  {York} B.,  {Hobbs} L.~M.,  {Welty} D.~E.,  {Friedman}
  S.~D.,  {Dahlstrom} J.,  {Snow} T.~P.,   {York} D.~G.,  2018, \mn@doi [\apjs]
  {10.3847/1538-4365/aad4a5}, \href
  {https://ui.adsabs.harvard.edu/abs/2018ApJS..237...40S} {237, 40}

\bibitem[\protect\citeauthoryear{{Spieler} et~al.,}{{Spieler}
  et~al.}{2017}]{2017ApJ...846..168S}
{Spieler} S.,  et~al., 2017, \mn@doi [\apj] {10.3847/1538-4357/aa82bc}, \href
  {http://adsabs.harvard.edu/abs/2017ApJ...846..168S} {846, 168}

\bibitem[\protect\citeauthoryear{{Strom}, {Strom}, {Carrasco}  \&
  {Vrba}}{{Strom} et~al.}{1975}]{1975ApJ...196..489S}
{Strom} K.~M.,  {Strom} S.~E.,  {Carrasco} L.,   {Vrba} F.~J.,  1975, \mn@doi
  [\apj] {10.1086/153429}, \href
  {https://ui-adsabs-harvard-edu.proxy1.lib.uwo.ca/abs/1975ApJ...196..489S}
  {196, 489}

\bibitem[\protect\citeauthoryear{{Thorburn} et~al.,}{{Thorburn}
  et~al.}{2003}]{C2DIB}
{Thorburn} J.~A.,  et~al., 2003, \mn@doi [\apj] {10.1086/345665}, \href
  {https://ui-adsabs-harvard-edu.proxy1.lib.uwo.ca/abs/2003ApJ...584..339T}
  {584, 339}

\bibitem[\protect\citeauthoryear{Van~der Maaten \& Hinton}{Van~der Maaten \&
  Hinton}{2008}]{van2008visualizing}
Van~der Maaten L.,  Hinton G.,  2008, Journal of machine learning research, 9

\bibitem[\protect\citeauthoryear{{Vos}, {Cox}, {Kaper}, {Spaans}  \&
  {Ehrenfreund}}{{Vos} et~al.}{2011}]{Vos2011}
{Vos} D.~A.~I.,  {Cox} N.~L.~J.,  {Kaper} L.,  {Spaans} M.,   {Ehrenfreund} P.,
   2011, \mn@doi [\aap] {10.1051/0004-6361/200809746}, \href
  {https://ui-adsabs-harvard-edu.proxy1.lib.uwo.ca/abs/2011A&A...533A.129V}
  {533, A129}

\bibitem[\protect\citeauthoryear{{Vuong} \& {Foing}}{{Vuong} \&
  {Foing}}{2000}]{2000A&A...363L...5V}
{Vuong} M.~H.,  {Foing} B.~H.,  2000, \aap, \href
  {https://ui-adsabs-harvard-edu.proxy1.lib.uwo.ca/abs/2000A&A...363L...5V}
  {363, L5}

\bibitem[\protect\citeauthoryear{{Walker}, {Webster}, {Bohlender}  \&
  {Kre{\l}owski}}{{Walker} et~al.}{2001}]{2001ApJ...561..272W}
{Walker} G. A.~H.,  {Webster} A.~S.,  {Bohlender} D.~A.,   {Kre{\l}owski} J.,
  2001, \mn@doi [\apj] {10.1086/323208}, \href
  {https://ui.adsabs.harvard.edu/abs/2001ApJ...561..272W} {561, 272}

\bibitem[\protect\citeauthoryear{{Walker}, {Bohlender}, {Maier}  \&
  {Campbell}}{{Walker} et~al.}{2015}]{2015ApJ...812L...8W}
{Walker} G.~A.~H.,  {Bohlender} D.~A.,  {Maier} J.~P.,   {Campbell} E.~K.,
  2015, \mn@doi [\apjl] {10.1088/2041-8205/812/1/L8}, \href
  {https://ui-adsabs-harvard-edu.proxy1.lib.uwo.ca/abs/2015ApJ...812L...8W}
  {812, L8}

\bibitem[\protect\citeauthoryear{{Walker}, {Campbell}, {Maier}, {Bohlender}  \&
  {Malo}}{{Walker} et~al.}{2016}]{2016ApJ...831..130W}
{Walker} G.~A.~H.,  {Campbell} E.~K.,  {Maier} J.~P.,  {Bohlender} D.,   {Malo}
  L.,  2016, \mn@doi [\apj] {10.3847/0004-637X/831/2/130}, \href
  {http://adsabs.harvard.edu/abs/2016ApJ...831..130W} {831, 130}

\bibitem[\protect\citeauthoryear{{Walker}, {Campbell}, {Maier}  \&
  {Bohlender}}{{Walker} et~al.}{2017}]{2017ApJ...843...56W}
{Walker} G.~A.~H.,  {Campbell} E.~K.,  {Maier} J.~P.,   {Bohlender} D.,  2017,
  \mn@doi [\apj] {10.3847/1538-4357/aa77f9}, \href
  {http://adsabs.harvard.edu/abs/2017ApJ...843...56W} {843, 56}

\bibitem[\protect\citeauthoryear{{Wampler}}{{Wampler}}{1966}]{1966ApJ...144..921W}
{Wampler} E.~J.,  1966, \mn@doi [\apj] {10.1086/148690}, \href
  {https://ui-adsabs-harvard-edu.proxy1.lib.uwo.ca/abs/1966ApJ...144..921W}
  {144, 921}

\bibitem[\protect\citeauthoryear{{Ward}}{{Ward}}{1963}]{WardMethod}
{Ward} J. H.~J.,  1963, \mn@doi [Journal of the American Statistical
  Association] {10.1080/01621459.1963.10500845}, 58, 236

\bibitem[\protect\citeauthoryear{{Webster}}{{Webster}}{1993}]{1993MNRAS.262..831W}
{Webster} A.,  1993, \mn@doi [\mnras] {10.1093/mnras/262.4.831}, \href
  {https://ui-adsabs-harvard-edu.proxy1.lib.uwo.ca/abs/1993MNRAS.262..831W}
  {262, 831}

\bibitem[\protect\citeauthoryear{{Welty} \& {Crowther}}{{Welty} \&
  {Crowther}}{2010}]{2010MNRAS.404.1321W}
{Welty} D.~E.,  {Crowther} P.~A.,  2010, \mn@doi [\mnras]
  {10.1111/j.1365-2966.2010.16386.x}, \href
  {https://ui-adsabs-harvard-edu.proxy1.lib.uwo.ca/abs/2010MNRAS.404.1321W}
  {404, 1321}

\bibitem[\protect\citeauthoryear{{Welty} \& {Hobbs}}{{Welty} \&
  {Hobbs}}{2001}]{2001ApJS..133..345W}
{Welty} D.~E.,  {Hobbs} L.~M.,  2001, \mn@doi [\apjs] {10.1086/320354}, \href
  {https://ui-adsabs-harvard-edu.proxy1.lib.uwo.ca/abs/2001ApJS..133..345W}
  {133, 345}

\bibitem[\protect\citeauthoryear{{Welty}, {Federman}, {Gredel}, {Thorburn}  \&
  {Lambert}}{{Welty} et~al.}{2006}]{2006ApJS..165..138W}
{Welty} D.~E.,  {Federman} S.~R.,  {Gredel} R.,  {Thorburn} J.~A.,   {Lambert}
  D.~L.,  2006, \mn@doi [\apjs] {10.1086/504153}, \href
  {https://ui-adsabs-harvard-edu.proxy1.lib.uwo.ca/abs/2006ApJS..165..138W}
  {165, 138}

\bibitem[\protect\citeauthoryear{{Welty}, {Ritchey}, {Dahlstrom}  \&
  {York}}{{Welty} et~al.}{2014}]{2014ApJ...792..106W}
{Welty} D.~E.,  {Ritchey} A.~M.,  {Dahlstrom} J.~A.,   {York} D.~G.,  2014,
  \mn@doi [\apj] {10.1088/0004-637X/792/2/106}, \href
  {https://ui-adsabs-harvard-edu.proxy1.lib.uwo.ca/abs/2014ApJ...792..106W}
  {792, 106}

\bibitem[\protect\citeauthoryear{{Wenger} et~al.,}{{Wenger}
  et~al.}{2000}]{2000A&AS..143....9W}
{Wenger} M.,  et~al., 2000, \mn@doi [\aaps] {10.1051/aas:2000332}, \href
  {https://ui.adsabs.harvard.edu/abs/2000A&AS..143....9W} {143, 9}

\bibitem[\protect\citeauthoryear{{Westerlund} \& {Krelowski}}{{Westerlund} \&
  {Krelowski}}{1989}]{1989A&A...218..216W}
{Westerlund} B.~E.,  {Krelowski} J.,  1989, \aap, \href
  {https://ui.adsabs.harvard.edu/abs/1989A&A...218..216W} {218, 216}

\bibitem[\protect\citeauthoryear{{Wszo{\l}ek} \& {God{\l}owski}}{{Wszo{\l}ek}
  \& {God{\l}owski}}{2003}]{2003MNRAS.338..990W}
{Wszo{\l}ek} B.,  {God{\l}owski} W.,  2003, \mn@doi [\mnras]
  {10.1046/j.1365-8711.2003.06143.x}, \href
  {https://ui-adsabs-harvard-edu.proxy1.lib.uwo.ca/abs/2003MNRAS.338..990W}
  {338, 990}

\bibitem[\protect\citeauthoryear{{Zhen}, {Castellanos}, {Paardekooper},
  {Linnartz}  \& {Tielens}}{{Zhen} et~al.}{2014}]{2014ApJ...797L..30Z}
{Zhen} J.,  {Castellanos} P.,  {Paardekooper} D.~M.,  {Linnartz} H.,
  {Tielens} A. G.~G.~M.,  2014, \mn@doi [\apjl] {10.1088/2041-8205/797/2/L30},
  \href
  {https://ui-adsabs-harvard-edu.proxy1.lib.uwo.ca/abs/2014ApJ...797L..30Z}
  {797, L30}

\bibitem[\protect\citeauthoryear{{Zuo}, {Li}  \& {Zhao}}{{Zuo}
  et~al.}{2021}]{2021ApJS..252...22Z}
{Zuo} W.,  {Li} A.,   {Zhao} G.,  2021, \mn@doi [\apjs]
  {10.3847/1538-4365/abcc6d}, \href
  {https://ui-adsabs-harvard-edu.proxy1.lib.uwo.ca/abs/2021ApJS..252...22Z}
  {252, 22}

\bibitem[\protect\citeauthoryear{{van Loon} et~al.,}{{van Loon}
  et~al.}{2013}]{2013A&A...550A.108V}
{van Loon} J.~T.,  et~al., 2013, \mn@doi [\aap] {10.1051/0004-6361/201220210},
  \href
  {https://ui-adsabs-harvard-edu.proxy1.lib.uwo.ca/abs/2013A&A...550A.108V}
  {550, A108}

\makeatother
\end{thebibliography}




\appendix

\section{Sight Line Information}
The following table contains basic information about the target stars and sight lines, along with their low-reddening standard stars used in the DIB measuring process. A full version of this table can be found in \cite{APOcat}.

\begin{table*}
	\centering
	\caption{Information on the 25 Target Sight Lines}
	\label{table: sightline info}
	\begin{threeparttable}
	\begin{tabular}{ccccccccc} 
		\hline
		\hline
		Star Name & Identifier & \makecell{Spectral Type} & $E_\textrm{B-V}$ & $f_{H2}$\tnote{1} & \makecell{Std. Star} & \makecell{Std. Star\\Spectral Type} & \makecell{Std. Star\\$E_\textrm{B-V}$} & Remarks \\
		(HD/BD) &  &  & (mag) &  & (HD) &  & (mag) &  \\
		\hline
		20041 &  & A0Ia & 0.72 & 0.42$^*$ & 46300 & A0Ib & 0.01 &  \\
		BD +40$^{\circ}$4220 & Cernis 52 & A3V & 0.90 & >0.78$^*$ & 107966 & A3V & 0.00 & \makecell{Report of PAH\\ \citep{2009ApJ...706..866G}} \\
		23180 & omi Per & B1III+B2V & 0.31 & 0.55 & 44743 & B1II-III & 0.02 & Steep ext. curve; Broad 2175 \AA bump. \\
		281159 &  & B5V & 0.85 & 0.50$^*$ & 16219 & B5V & 0.04 & \\
		23512 &  & A0V & 0.36 & 0.62$^*$ & 31647 & A1V & 0.01 & Steep ext. curve; Broad 2175 \AA bump.\\
		24534 & X Per & O9.5pe & 0.59 & 0.76 & 214680 & O9V & 0.11 & \makecell{Translucent cloud; Steep ext.curve;\\ Broad 2175 \AA bump.}\\
		24912 & xi Per & O7e & 0.33 & 0.38 & 47839 & O7Ve & 0.07 & \\
		28482 &  & B8III & 0.52 & 0.66$^*$ & 4382 & B8III & 0.01 & Steep ext. curve; Broad 2175 \AA bump. \\
		37061 & NU Ori & B1V & 0.52 & 0.02$^*$ & 36959 & B1V & 0.03 & \makecell{Intense radiation field; Flat ext. curve;\\ Weak 2175 \AA bump.} \\
		37903 &  & B1.5V & 0.35 & 0.53 & 37018 & B1V & 0.07 &\makecell{Anamalous 5780/5797 ratio; Flat ext.\\curve; Weak 2175 \AA bump.} \\
		43384 & 9 Gem & B3Ib & 0.58 & 0.44$^*$ & 52089 & B2II & 0.01 & \\
		147084 & omi Sco & A5II & 0.73 & 0.59$^*$ & 186377 & A5III & 0.04 & \\
		147889 &  & B2V & 1.07 & 0.45 & 42690 & B2V & 0.04 & \makecell{Embedded and ionizing nearby cloud\\ \citep{2013MNRAS.428.2617R}; Steep ext. curve.} \\
		148579 &  & B9V & 0.34 & 0.45$^*$ & 201433 & B9V & 0.00 & Flat ext. curve; Weak 2175 \AA bump.\\
		166734 &  & O8e & 1.39 & 0.39$^*$ & 47839 & O7Ve & 0.07 & \\
		168625 &  & B8Ia & 1.48 & 0.33$^*$ & 34085 & B8Iae & 0.00 & \\
		175156 &  & B5II & 0.31 & 0.31$^*$ & 34503 & B5III & 0.05 & Steep ext. curve; Weak 2175 \AA bump. \\
		183143 &  & B7Iae & 1.27 & 0.31$^*$ & 63975 & B8II & 0.00 & \\
		190603 &  & B1.5Iae & 0.72 & 0.16 & 52089 & B2II & 0.01 & Flat ext. curve; Weak 2175 \AA bump. \\
		194279 &  & B2Iae & 1.20 & 0.30$^*$ & 53138 & B3Iab & 0.05 & \makecell{Multiple components but average\\ condition \citep{Cox2011}} \\
		BD +40$^{\circ}$4220 & VI Cyg 5 & O7f & 1.99 & 0.47$^*$ & 47839 & O7Ve & 0.07 & \\
		 & VI Cyg 12 & B5Ie & 1.11 & 0.67$^*$ & 36959 & B1V & 0.03 & Schulte's Star.\\
		204827 &  & \makecell{O9.5V+\\B0.5III} & 1.11 & 0.67$^*$ & 36959 & B1V & 0.03 & Steep ext. Curve; Weak 2175 \AA bump. \\
		206267 &  & O6f & 0.53 & 0.42 & 47839 & O7Ve & 0.07 & \\
		223385 & 6Cas & A3Iae & 0.67 & 0.12$^*$ & 197345 & A2Ia & 0.09 & \\
		\hline
	\end{tabular}
  \begin{tablenotes}
    \item[1] Values marked by asterisks are surrogate result, i.e. $N$(H) estimated from $W$(5780) and/or $N$(H$_2$) estimated from $N$(CH). See \cite{behavior} for details.
  \end{tablenotes}
  \end{threeparttable}
\end{table*}

\section{Page-Wide Plots}\label{appendix: page-wide plots}
\begin{figure*}
	\includegraphics[width=\textwidth]{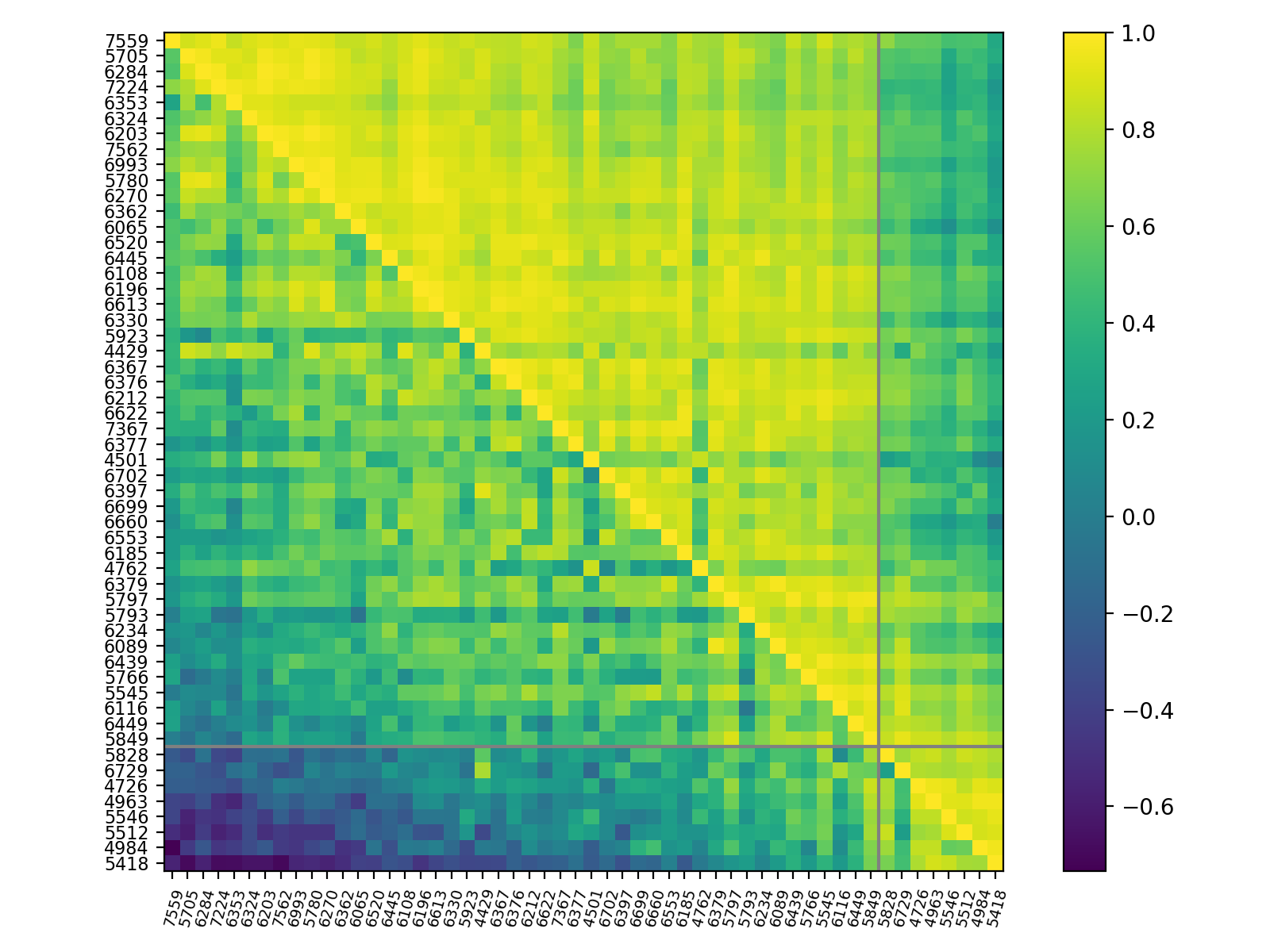}
    \caption{Heat map for the Pearson correlation coefficients ($r$ values) among all 54 target DIBs. The figure is asymmetrical, where the lower triangle is for the $r_\textrm{norm}$ values and the upper triangle is for the $r_\textrm{reg}$ values. The DIBs are sorted by the sequence in Section \ref{sec:MDS}, and the grey lines separate the C$_2$ DIBs from the non-C$_2$ DIBs. The transition in the colours and hence $r$ values is rather smooth, except around the grey lines, which is more obvious in the $r_\textrm{reg}$ section.}
    \label{fig: big_heat_map}
\end{figure*}

\begin{figure*}
	\includegraphics[width=\textwidth]{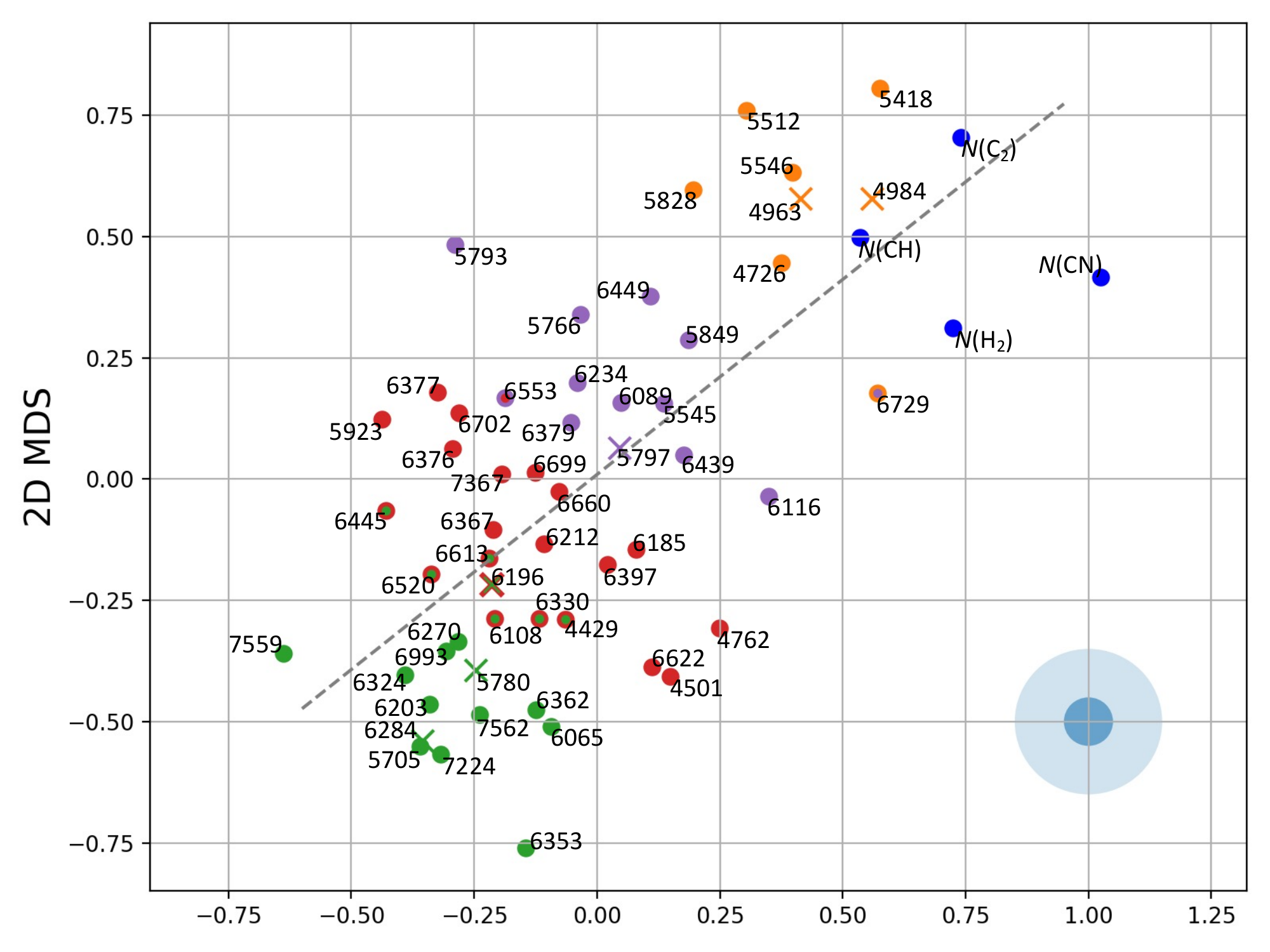}
    \caption{Result of 2D-MDS where all data points are labelled. Colours reflect the clustering result as in Figure \ref{fig: Dendrogram}, where green is for the $\sigma$-type DIBs, red is for the intermediate DIBs, purple is for the $\zeta$-type DIBs, orange is for the C$_2$ DIBs, and blue for the column densities of molecular species included for comparison. The ten DIBs assigned to different groups by HAC and $k$-means algorithms are plotted with different colours, where the face (inner) colour represents the HAC result and the edge (outer) colour represents the $k$-means result. DIBs $\lambda\lambda$6284, 5780, 6196, 5797, 4963, and 4984 are highlighted as ``X'' for their known sequence. The transparent blue circles indicate 0.05 and 0.15 radius on the plot, and the dashed line is the best-fit line of all DIB points. It follows the sequence of DIBs and acts as the the first projection axis. We find a smooth and continuous distribution among the non-C$_2$ DIBs, whereas the C$_2$ DIBs seem to form a separate cluster with other small molecules.}
    \label{fig: MDS_result_all_label}
\end{figure*}

This appendix contains page-wide figures to include more details and information than their counterparts in the main text. Figure \ref{fig: big_heat_map} demonstrates $r_\textrm{reg}$ (upper triangle) and $r_\textrm{norm}$ (lower triangle) values among the 54 target DIBs. The DIBs are sorted by the sequence proposed in Section \ref{sec:MDS}. As in the MDS result, the transitions in the $r$ values (colours) is smooth across the DIBs, except around the grey lines separating the C$_2$ DIBs from the non-C$_2$ DIBs. This gap is more obvious in the $r_{reg}$ values. Figure \ref{fig: MDS_result_all_label} is the enlarged version of Figure \ref{fig: MDS_result}, except all data points are labelled. Note the colours represents the clustering results (Section \ref{sec:clustering}) and DIBs assigned to different groups by HAC and $k$-means clustering algorithms are plotted in two colours. Detailed discussions on this plot can be found in Sections \ref{sec:MDS} and \ref{sec:factors}.

\section{Analysis Using $r_{\rm reg}$ Values}
\label{appendix:different_r}
We focus on the $r_\textrm{norm}$ values in the main text since this work is originally motivated by the search of anti-correlated DIBs. But in principle, the clustering and MDS analyses can be applied to the $r_\textrm{reg}$ matrix as well. Here we follow the same routine as in Sections \ref{sec:clustering} and \ref{sec:MDS} to show using $r_\textrm{reg}$ would not change most of our findings.

\begin{figure*}
	\includegraphics[width=\textwidth]{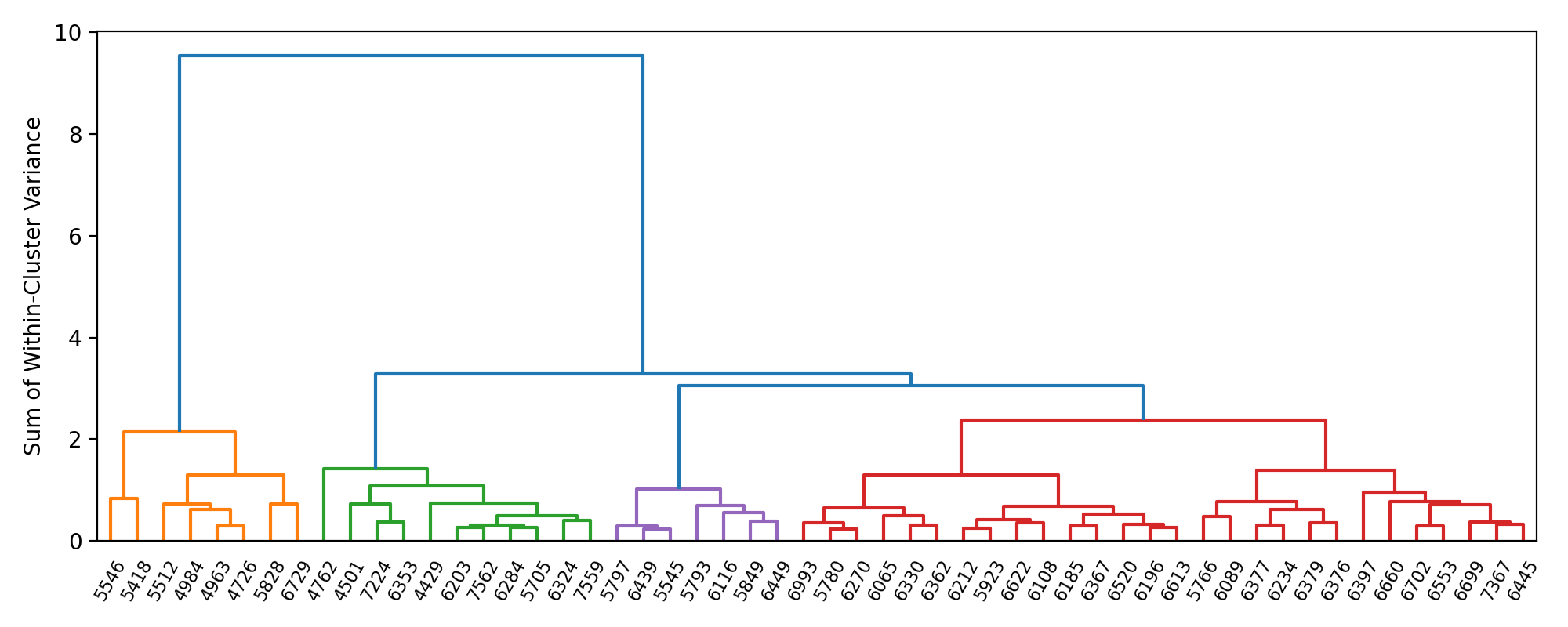}
    \caption{Dendrogram using the $r_\textrm{reg}$ values. DIBs connected by a lower horizontal bar (node) share more similarities. We note the difference between C$_2$ and non-C$_2$ DIBs is more obvious, and the assignments among the non-C$_2$ DIBs can be different compared to Figure \ref{fig: Dendrogram}.}
    \label{fig: Dendrogram_r_regular}
\end{figure*}

We start with sorting the target DIBs into four clusters using the $r_\textrm{reg}$ value matrix. Figure \ref{fig: Dendrogram_r_regular} shows the dendrogram from HAC, and the difference between C$_2$ and non-C$_2$ DIBs is more highlighted. The assignments among the three non-C$_2$ groups are somewhat shuffled. For example, the intermediate group (red) now contains about half of the non-C$_2$ DIBs and includes $\lambda$5780, which is traditionally recognized as a $\sigma$-type DIB. In Table \ref{table: 4way clustering} we summarize the clustering results from different methods and inputs (i.e. the HAC or $k$-means clustering algorithms, $r_\textrm{norm}$ or $r_\textrm{reg}$ value matrix). The DIBs are ordered by the sequence from Section \ref{sec:MDS}. For all combinations of method and input data, the assignments of DIBs generally follow the flow as $\sigma$-type, intermediate, $\zeta$-type, and C$_2$ DIBs, and the boundary between C$_2$ and non-C$_2$ DIBs is quite solid. We find 34 out of 54 target DIBs being assigned to the same group throughout different methods and inputs, and the rest are assigned to two groups. There is no DIB assigned to three different groups in our analysis.

\begin{table*}
	\centering
	\caption{Results of Different Clustering Algorithms and Inputs}
	\label{table: 4way clustering}
	\begin{threeparttable}
	\begin{tabular}{cccccc|cccccc} 
		\hline
		\multirow{2}{*}{Idx\tnote{1}} & \multirow{2}{*}{DIB} & \multicolumn{2}{c}{$r_\textrm{norm}$} & \multicolumn{2}{c|}{$r_\textrm{reg}$} & \multirow{2}{*}{Idx\tnote{1}} & \multirow{2}{*}{DIB} & \multicolumn{2}{c}{$r_\textrm{norm}$} & \multicolumn{2}{c}{$r_\textrm{reg}$} \\
		 &  & HAC & $k$-means & HAC & $k$-means &  &  & HAC & $k$-means & HAC & $k$-means \\
		\hline
		1 & 7559 & $\sigma$ & $\sigma$ & $\sigma$ & $\sigma$ & 28 & 4501 & Inter. & Inter. & $\sigma$ & $\sigma$ \\
        2 & 5705 & $\sigma$ & $\sigma$ & $\sigma$ & $\sigma$ & 29 & 6702 & Inter. & Inter. & Inter. & Inter. \\
        3 & 6284 & $\sigma$ & $\sigma$ & $\sigma$ & $\sigma$ & 30 & 6397 & Inter. & Inter. & Inter. & Inter. \\
        4 & 7224 & $\sigma$ & $\sigma$ & $\sigma$ & $\sigma$ & 31 & 6699 & Inter. & Inter. & Inter. & Inter. \\
        5 & 6353 & $\sigma$ & $\sigma$ & $\sigma$ & $\sigma$ & 32 & 6660 & Inter. & Inter. & Inter. & Inter. \\
        6 & 6324 & $\sigma$ & $\sigma$ & $\sigma$ & $\sigma$ & 33 & 6553 & Inter. & $\zeta$ & Inter. & Inter. \\
        7 & 6203 & $\sigma$ & $\sigma$ & $\sigma$ & $\sigma$ & 34 & 6185 & Inter. & Inter. & Inter. & Inter. \\
        8 & 7562 & $\sigma$ & $\sigma$ & $\sigma$ & $\sigma$ & 35 & 4762 & Inter. & Inter. & $\sigma$ & $\sigma$ \\
        9 & 6993 & $\sigma$ & $\sigma$ & Inter. & $\sigma$ & 36 & 6379 & $\zeta$ & $\zeta$ & Inter. & Inter. \\
        10 & 5780 & $\sigma$ & $\sigma$ & Inter. & $\sigma$ & 37 & 5797 & $\zeta$ & $\zeta$ & $\zeta$ & $\zeta$ \\
        11 & 6270 & $\sigma$ & $\sigma$ & Inter. & Inter. & 38 & 5793 & $\zeta$ & $\zeta$ & $\zeta$ & $\zeta$ \\
        12 & 6362 & $\sigma$ & $\sigma$ & Inter. & Inter. & 39 & 6234 & $\zeta$ & $\zeta$ & Inter. & Inter. \\
        13 & 6065 & $\sigma$ & $\sigma$ & Inter. & Inter. & 40 & 6089 & $\zeta$ & $\zeta$ & Inter. & $\zeta$ \\
        14 & 6520 & $\sigma$ & Inter. & Inter. & Inter. & 41 & 6439 & $\zeta$ & $\zeta$ & $\zeta$ & $\zeta$ \\
        15 & 6445 & $\sigma$ & Inter. & Inter. & Inter. & 42 & 5766 & $\zeta$ & $\zeta$ & Inter. & $\zeta$ \\
        16 & 6108 & $\sigma$ & Inter. & Inter. & Inter. & 43 & 5545 & $\zeta$ & $\zeta$ & $\zeta$ & $\zeta$ \\
        17 & 6196 & $\sigma$ & Inter. & Inter. & Inter. & 44 & 6116 & $\zeta$ & $\zeta$ & $\zeta$ & $\zeta$ \\
        18 & 6613 & $\sigma$ & Inter. & Inter. & Inter. & 45 & 6449 & $\zeta$ & $\zeta$ & $\zeta$ & $\zeta$ \\
        19 & 6330 & $\sigma$ & Inter. & Inter. & Inter. & 46 & 5849 & $\zeta$ & $\zeta$ & $\zeta$ & $\zeta$ \\
        20 & 5923 & Inter. & Inter. & Inter. & Inter. & 47 & 5828 & C$_2$ & C$_2$ & C$_2$ & C$_2$ \\
        21 & 4429 & $\sigma$ & Inter. & $\sigma$ & $\sigma$ & 48 & 6729 & $\zeta$ & C$_2$ & C$_2$ & C$_2$ \\
        22 & 6367 & Inter. & Inter. & Inter. & Inter. & 49 & 4726 & C$_2$ & C$_2$ & C$_2$ & C$_2$ \\
        23 & 6376 & Inter. & Inter. & Inter. & Inter. & 50 & 4963 & C$_2$ & C$_2$ & C$_2$ & C$_2$ \\
        24 & 6212 & Inter. & Inter. & Inter. & Inter. & 51 & 5546 & C$_2$ & C$_2$ & C$_2$ & C$_2$ \\
        25 & 6622 & Inter. & Inter. & Inter. & Inter. & 52 & 5512 & C$_2$ & C$_2$ & C$_2$ & C$_2$ \\
        26 & 7367 & Inter. & Inter. & Inter. & Inter. & 53 & 4984 & C$_2$ & C$_2$ & C$_2$ & C$_2$ \\
        27 & 6377 & Inter. & Inter. & Inter. & Inter. & 54 & 5418 & C$_2$ & C$_2$ & C$_2$ & C$_2$ \\
		\hline
	\end{tabular}
  \begin{tablenotes}
    \item[1] DIBs are sorted according to the sequence in Section \ref{sec:MDS}
  \end{tablenotes}
  \end{threeparttable}
\end{table*}

\begin{figure}
	\includegraphics[width=\columnwidth]{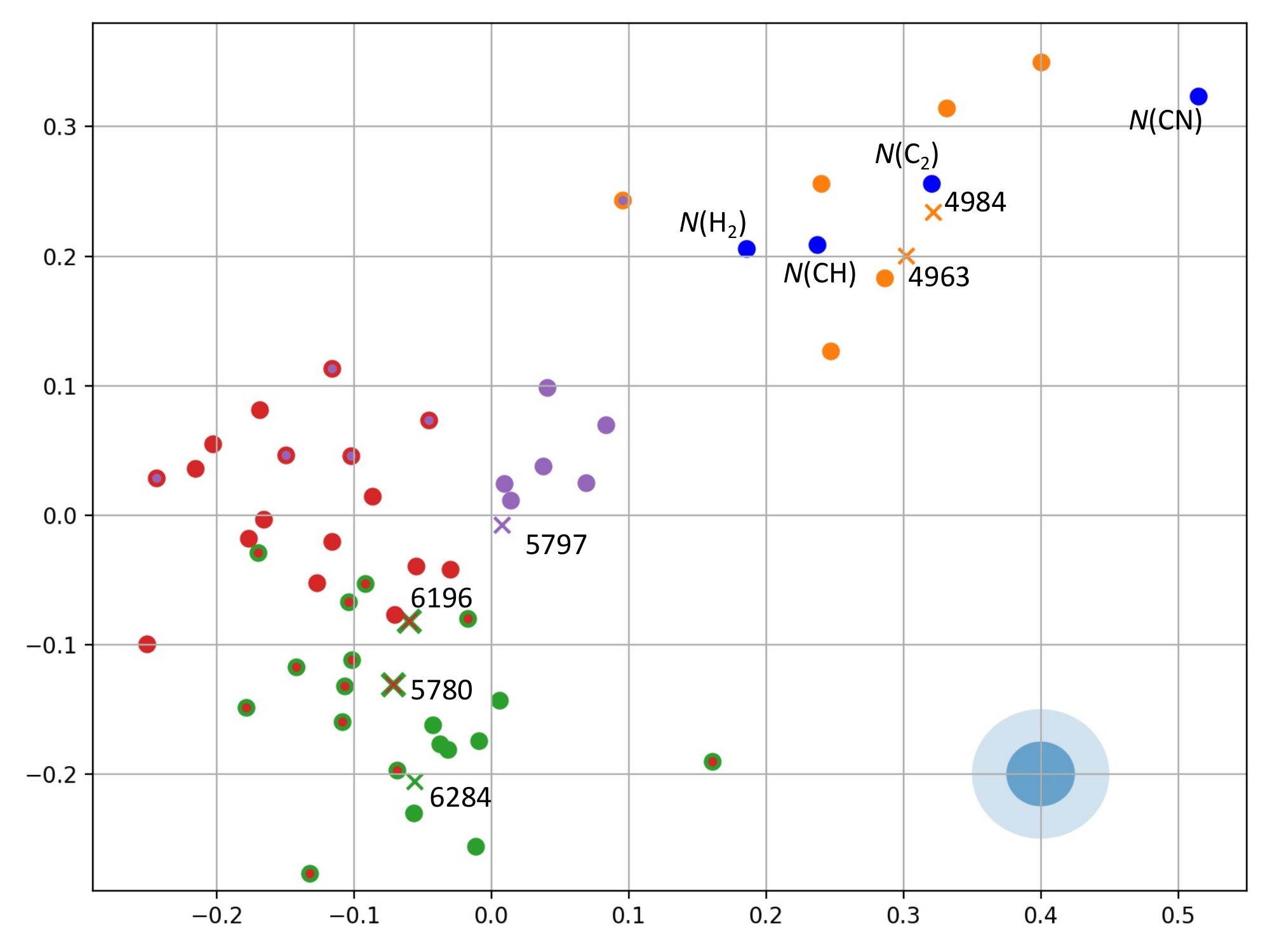}
    \caption{2D-MDS result where distance between points represents 1 - $r_\textrm{reg}$ dissimilarity. Colours represent memberships of group, i.e. green for the $\sigma$-type, red for intermediate, purple for the $\zeta$-type, orange for the C$_2$ DIBs, and blue for known molecular species. DIBs assigned to two groups in Table \ref{table: 4way clustering} have distinct face and edge colours. DIBs $\lambda\lambda$6284, 5780, 6196, 5797, 4963, and 4984 are highlighted as ``X'' for their known sequence. The transparent blue circles at the lower-right corner indicate 0.05 and 0.1 radius in the plot. The C$_2$ DIBs remain clustered with molecular species and are more separated from the non-C$_2$ DIBs compared to Figure \ref{fig: MDS_result}. We find the overall trend bends into an arch, which might be due to the non-linear effects in correlation coefficient.}
    \label{fig: MDS_r_regular}
\end{figure}

Figure \ref{fig: MDS_r_regular} presents the 2D-MDS results from the 1 - $r_\textrm{reg}$ matrix. Compared to Figure \ref{fig: MDS_result} the C$_2$ DIBs here are still clustered with the molecular species and more separated from the non-C$_2$ DIBs. The continuous distribution among non-C$_2$ DIBs remain valid, and there is no clear boundary to further divide them into smaller clusters. On the other hand, the overall trend bends into an arch. It highlights the necessity of including a second projection axis, and this arched distribution might be related to the non-linear effect in correlation coefficients. 

To sum-up, analyses based on the $r_\textrm{reg}$ values would provide a similar picture on DIB correlations and strengthens the discussions we made in the text, especially regarding the following:
\begin{enumerate}
    \item The C$_2$ DIBs are a unique DIB family that are more clustered with molecular species and are well separated from the non-C$_2$ DIBs.
    \item There is a rather continuous transitions among the non-C$_2$ DIBs, although some of the members may demonstrate quite different behaviour. This continuous trend is against further sorting all non-C$_2$ DIBs into clearly distinguishable groups, and terms like $\sigma$- or $\zeta$-type DIBs may only be applied to the most representing members.
    \item At least two factors are needed to properly reproduce the DIB correlation.
\end{enumerate}


\bsp	
\label{lastpage}

\end{document}